\documentclass[a4paper,twocolumn,11pt,unpublished]{quantumarticle}
\pdfoutput=1

\usepackage[utf8]{inputenc}
\usepackage[T1]{fontenc}

\usepackage[english]{babel}
\usepackage{graphicx}
\usepackage{xcolor}
\usepackage{bm}
\usepackage{braket}
\usepackage{dcolumn}
\usepackage{multirow}
\usepackage{quantikz}
\usetikzlibrary{quantikz2}
\usepackage{stackengine}
\usepackage[numbers,sort]{natbib}

\usepackage[colorlinks=true,citecolor=blue,linkcolor=blue,urlcolor=blue]{hyperref}

\begin{document}

\title{Circuit-Level Loss Performance of RHG and Foliated Floquet Color Codes in a Compound Photon--Atom Quantum Architecture}

\author{Dana Ben Porath}
\author{Juval Bechar}
\author{Daniel Azses}
\author{Yaron Jarach}
\email{yaron@qs-labs.com}
\affiliation{Quantum Source Labs, Ness Ziona 7403636, Israel}

\begin{abstract}
A central question for fault-tolerant quantum computing is which quantum error-correcting codes are best suited to a given hardware architecture. Here we compare the Raussendorf--Harrington--Goyal (RHG) code, the Foliated
Floquet Color Code (FFCC), and the reduced FFCC in a compound photon--atom
architecture that directly generates measurement-based quantum
computation (MBQC) resources with near-deterministic photon--atom CZ
gates. RHG serves as a natural benchmark, while the FFCC variants allow
us to study whether reduced graph degree improves performance under an
architecture-aware circuit-level loss model with delayed heralding and
correlated bond-loss propagation. We construct two generation schemes
compatible with the compound hardware and evaluate circuit-level
thresholds under periodic boundary conditions. RHG achieves the highest
circuit-level threshold, 2.75\%, and its threshold falls below that of
reduced FFCC only for large excess loss on intermodule CZ connections.
RHG also achieves the lowest logical error rate in most resource-matched
comparisons, but some low-loss windows favor reduced FFCC. Overall, we
show that when the hardware supports the native gates and connectivity
required for MBQC, the benefits of lower graph degree must be weighed
against each code's intrinsic IID loss tolerance, generation-scheme
details, and hardware-aware resource overhead.
\end{abstract}

\maketitle

\section{Introduction}

Measurement-based quantum computation (MBQC) proceeds by preparing an entangled resource state, typically a cluster state, and consuming it through single-qubit measurements~\cite{raussendorf2001one,briegel2001persistent,raussendorf2003measurement,Dawson_2006, Briegel_2009}. It is particularly well suited to photonic architectures, where difficult-to-store flying qubits can be generated, entangled, routed, and measured as part of a growing resource state~\cite{browne2005resource,bourassa2021blueprint,bartolucci2023fusion}. In loss-dominated settings, MBQC can achieve higher thresholds than circuit-based constructions~\cite{Whiteside_2014,Perrin_2025,Baranes_2026,Kobayashi_2026}.

For quantum error correction, foliation provides a measurement-based
representation of repeated syndrome extraction~\cite{bolt2016prl}.
Successive syndrome-measurement rounds of a circuit-based code are
mapped to successive layers of a three-dimensional cluster state.
As each layer is measured, the encoded logical information is
teleported forward, while products of measurement outcomes provide
the syndrome information required for decoding. In this picture, \emph{data} qubits carry the propagated logical information,
while \emph{syndrome-like} (SL) qubits complete the error-correction
checks~\cite{raussendorf2007topological}.

In the compound photon--atom architecture studied here, stationary
atomic qubits and flying photonic qubits play complementary roles and
jointly carry quantum information~\cite{qs2026arxiv}. The basic
hardware element is a reusable atom--cavity unit cell that can generate
single photons, implement near-deterministic photon--atom CZ gates, and
perform atomic state preparation and measurement~\cite{qs2026arxiv,
aqua2025,duan_kimble_cz}. Reconfigurable optical routing directs photons
between these unit cells, so that a photon can be generated, interact
sequentially with selected atoms through CZ gates, and then be
measured. In this way, photons provide long-range connectivity across
the system, while atoms act as reusable entanglement sites and
short-term quantum memories. The resulting long-range connectivity enables flexible generation of the MBQC resource states, including the wraparound connections required here for the periodic boundary conditions.

This setting differs from purely photonic architectures, where large resource states are typically assembled using probabilistic linear-optical operations~\cite{browne2005resource,bartolucci2023fusion,bourassa2021blueprint}. It also differs from many matter-based architectures, where connectivity relies on local interactions or qubit transport~\cite{google2023surfacecode,bluvstein2024logical,mohseni2024build,megrant2025scaling,wang2026demonstration,bravyi2024high,mathews2026placing,landsman2019pra,johnson2025nonlinear,evered2023high,reichardt2024fault,khan2026architecting}. In other hybrid light--matter schemes, matter qubits primarily generate photons or finite-size spin--photon resources, while larger clusters are assembled through probabilistic fusion~\cite{chan2023onchip,meng2024deterministic,degliniasty2024spinoptical,simmons2024siliconcolour,chan2026practical}.

\begin{figure*}[!t]
\centering

\begin{minipage}[t]{0.45\textwidth}
    \centering
    \vspace{0pt}
    \stackinset{l}{0pt}{t}{2pt}{\textbf{(a)}}{%
        \parbox[c][6cm][c]{\linewidth}{%
            \centering
            \includegraphics[width=0.9\linewidth]{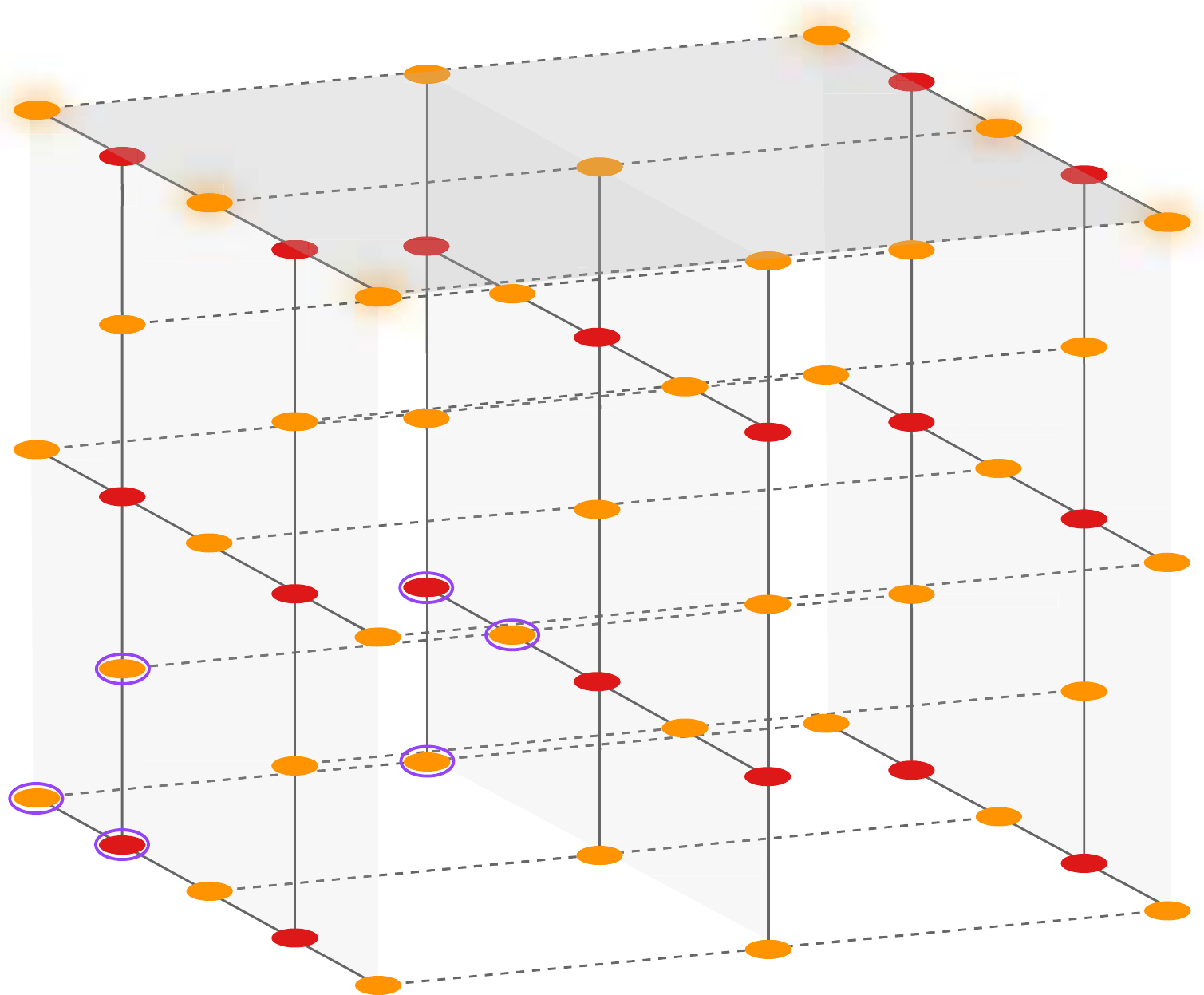}%
        }%
    }%
\end{minipage}
\hfill
\begin{minipage}[t]{0.54\textwidth}
    \centering
    \vspace{0pt}
    \stackinset{l}{0pt}{t}{2pt}{\textbf{(b)}}{%
        \parbox[c][6cm][c]{\linewidth}{%
            \centering
            \includegraphics[width=0.9\linewidth]{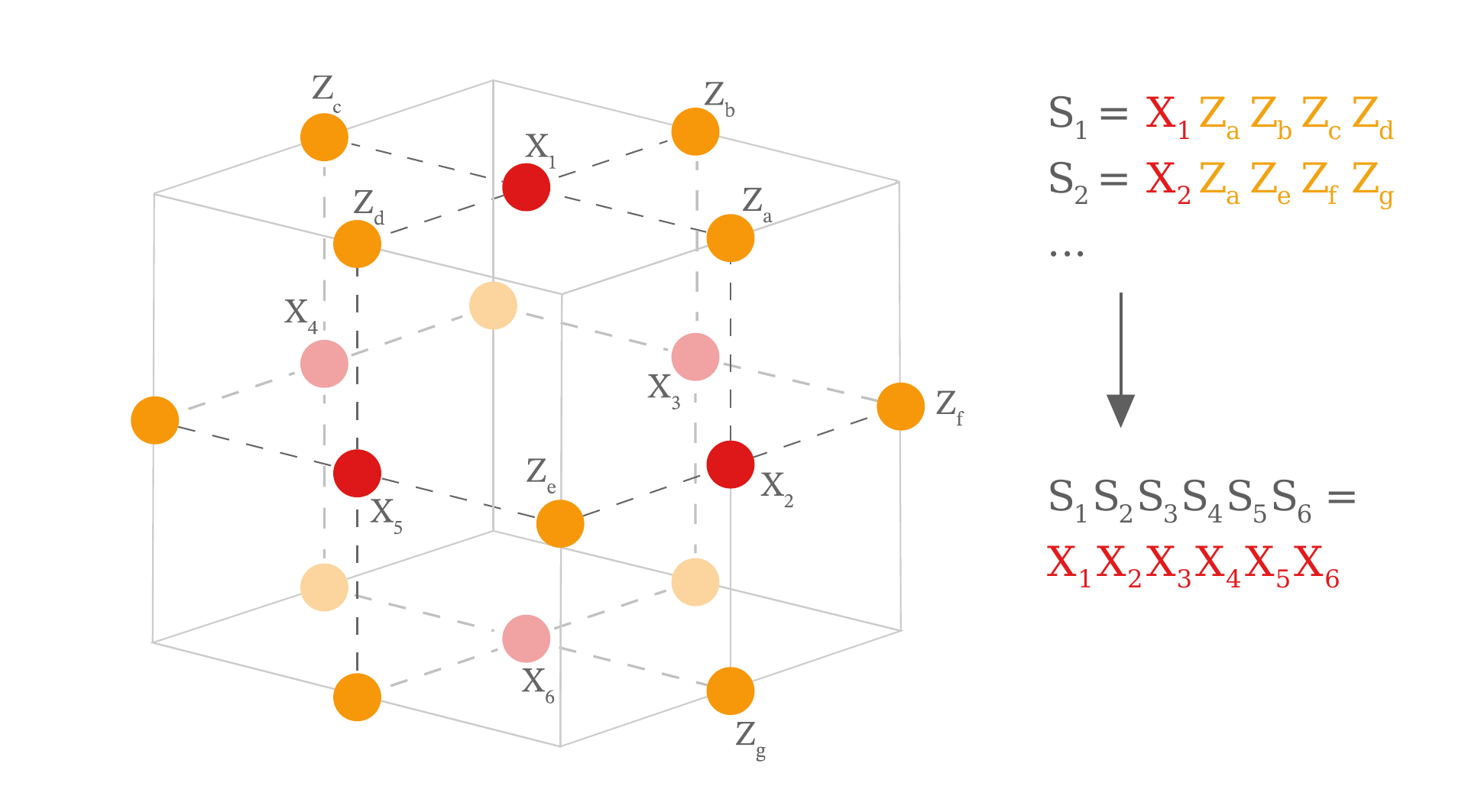}%
        }%
    }%
\end{minipage}
    \caption{
    RHG code construction and check structure, adapted from Arwas et al.~\cite{qs2026arxiv}.
    (a) An \(l=3\), \(t=3\) RHG cluster-state construction starting and ending with dual layers, corresponding to logical \(\ket{0}_{L}\) initialization and measurement. 
    For clarity, the wraparound CZ links implementing periodic boundaries in the simulations are omitted.
    Orange and red circles denote data and syndrome-like (SL) qubits, respectively; dashed and solid gray lines denote time-like and space-like CZ gates. The shaded cap marks a logical correlation surface determined by the product of the highlighted \(X\)-basis measurement outcomes. The circled qubits indicate a possible six-ring unit cell, motivating the distinction between intramodule and intermodule CZ loss used in Fig.~\ref{fig:intermodule_loss_comparison}.
    (b) An RHG code check obtained from a product of cluster-state stabilizers over a membrane enclosing a cube.
    }
    \label{fig:rhg}
\end{figure*}

\begin{figure*}[!t]
\centering

\makebox[0.48\textwidth][l]{%
    \stackinset{l}{0pt}{t}{2pt}{\textbf{(a)}}{%
        \hspace{18pt}%
        \includegraphics[width=0.42\textwidth]{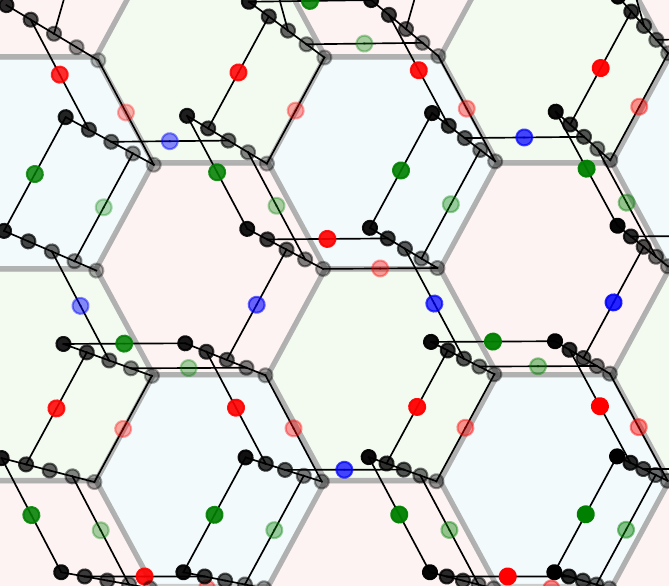}%
    }%
}%
\hfill
\makebox[0.48\textwidth][l]{%
    \stackinset{l}{0pt}{t}{2pt}{\textbf{(b)}}{%
        \hspace{18pt}%
        \includegraphics[width=0.42\textwidth]{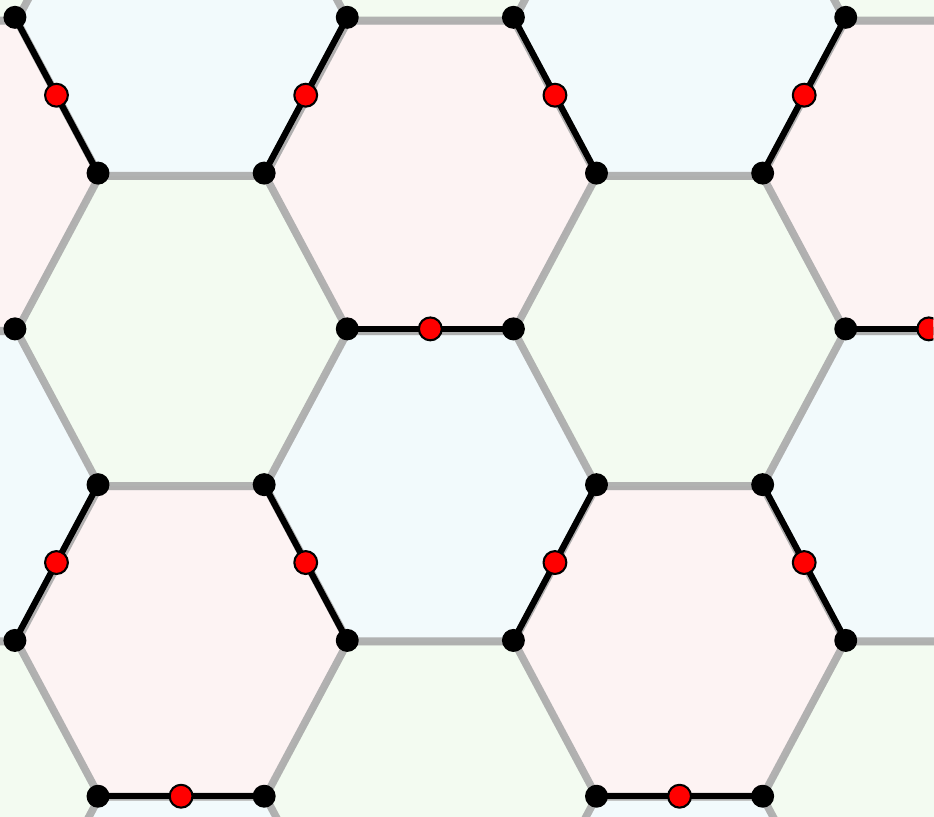}%
    }%
}%

\caption{
The FFCC code construction for $l=3$ with periodic boundary conditions on a rectangular cut. Black circles denote data qubits, colored circles denote SL qubits, and edges denote CZ gates. Edges through the periodic boundaries are omitted for clarity.
(a) Five layers ordered as red--green--blue--red--green.
(b) A single red layer. }
\label{fig:ffcc_cluster_layers}
\end{figure*}

We compare three MBQC-based codes: the
Raussendorf--Harrington--Goyal (RHG) code
(Fig.~\ref{fig:rhg}), the
Foliated Floquet Color Code (FFCC)
(Fig.~\ref{fig:ffcc_cluster_layers}), and the
reduced FFCC
(Fig.~\ref{fig:reduced_ffcc_structure}), abbreviated as rFFCC in figures and equations. RHG is the canonical three-dimensional cluster-state realization of the surface code and provides a well-established benchmark~\cite{raussendorf2006aop,
raussendorf2007topological}. FFCC is another foliated MBQC code, inspired by Floquet color codes, whose resource graph has lower degree than RHG~\cite{bolt2016prl,paesani2023prl}. Reduced FFCC further
removes the SL qubits and replaces them with local Hadamard operations,
yielding a more compact construction~\cite{paesani2023prl}. These codes
therefore provide a natural setting for studying whether reducing graph
degree improves loss performance. In our architecture, each graph
connection requires a CZ gate and thus introduces another opportunity
for loss. Lower degree might therefore seem advantageous. However, it
also changes the code geometry, checks, logical operators, qubit count,
generation scheme, and error propagation. We therefore ask which code
performs best when these effects are considered together. 

To compare the codes, we use a simplified architecture-aware circuit-level loss model described in Sec.~\ref{section:generation_schemes}. Importantly, the model includes photon loss during photon--atom CZ gates. Mid-circuit photon loss can remove subsequent bonds~\cite{li2010fault,auger2018fault} and induce correlated Pauli errors on neighboring qubits~\cite{yu2025processingdecodingrydbergdecay, qs2026arxiv}. Because loss is heralded only at measurement, its precise time is unknown. We condition the loss model on the known gate order and final loss outcome~\cite{gu2024optimizingquantumerrorcorrection,Baranes_2026,qs2026arxiv}. The resulting detector error model is sampled with \texttt{Stim} and decoded by minimum-weight perfect matching using PyMatching~\cite{gidney2021stim,higgott2021pymatchingpythonpackagedecoding,Higgott2025sparseblossom}.

Under an independent and identically distributed (IID) loss model, the thresholds are approximately 25\% for RHG~\cite{Barrett_2010}, 8.5\% for FFCC, and 13.5\% for reduced FFCC~\cite{paesani2023prl}. These benchmarks do not determine the best code for a given architecture because circuit-level performance depends on graph degree and gate order. We therefore compare two hardware-compatible generation schemes. In the bipartite scheme, each qubit remains either photonic or atomic throughout its entangling sequence. In the STAP (State Transfer from Atom to Photon) scheme, a qubit may be transferred from an atom to a photon between entangling gates~\cite{qs2026arxiv}. Comparing the two separates code properties from implementation-specific effects.

Several previous studies compared related low-degree codes with surface-code constructions under circuit-level Pauli or erasure noise. Gidney et al.~\cite{gidney2021honeycomb} compared honeycomb- and rotated-surface-code memories in unitary-gate circuit models and also analyzed a honeycomb implementation with native two-qubit measurements. Gu et al.~\cite{gu2025faulttolerantarchitectures,
gu2024optimizingquantumerrorcorrection} studied erasure-qubit
protocols under a different propagation model. Their baseline models assume that an erased qubit fully depolarizes the other qubit in each subsequent entangling operation. This erasure-depolarization model makes the degree dependence more direct and more Pauli-like than in the structured bond-loss model considered here.

Dessertaine et al.~\cite{dessertaine2026enhancedfaulttolerancephotonicquantum}
compared the honeycomb and surface codes in a hybrid spin--optical architecture. The proposed hybrid architecture combined spin-qubit photon emitters with probabilistic, repeat-until-success fusion measurements. The analysis included photon loss at measurement, spin-qubit decoherence, and photon indistinguishability. Related works studied the FFCC under similar assumptions~\cite{paesani2023prl,chan2025prxquantum}. These works found that the honeycomb and FFCC constructions outperform the corresponding surface-code construction~\cite{bartolucci2023fusion} under their fusion-based noise models. None considered direct generation of the MBQC resource through
near-deterministic native entangling gates.

Here, we study a distinct loss regime in directly generated MBQC. We construct periodic RHG, FFCC, and reduced FFCC resources using the same near-deterministic photon--atom CZ primitive and a common hardware-aware circuit-level loss model. We provide explicit hardware-compatible generation schemes and compare the resulting thresholds and subthreshold scaling of the three codes. We also model excess loss on intermodule CZ connections. Under uniform CZ loss, periodic RHG achieves the highest threshold among the three codes, whereas sufficiently large intermodule loss reverses the ordering in favor of reduced FFCC. This reversal shows that lower graph degree alone does not determine performance: gate ordering and the location of loss within the generation circuit also play an important role. More broadly, it demonstrates that code rankings depend on the native entangling primitive and its associated loss-propagation mechanism. This distinguishes our direct photon--atom CZ setting from previous photonic and hybrid schemes based on probabilistic fusion.

The work is organized as follows: In Section~\ref{section:code_construction}, we introduce the codes considered in this work. In Section~\ref{section:generation_schemes}, we describe the generation schemes used to implement them in the compound photon--atom architecture. In Section~\ref{section:results}, we present the resulting circuit-level loss thresholds, analyze logical-error-rate scaling, and compare the logical error rates (LERs) of the three codes under
matched resource budgets.

\begin{figure*}[!t]
\centering
\begin{minipage}[t]{0.3\textwidth}
    \centering
    \stackinset{l}{2pt}{t}{2pt}{\textbf{(a)}}{%
        \parbox[c][5.3cm][c]{\linewidth}{%
            \centering
            \includegraphics[width=0.95\linewidth]{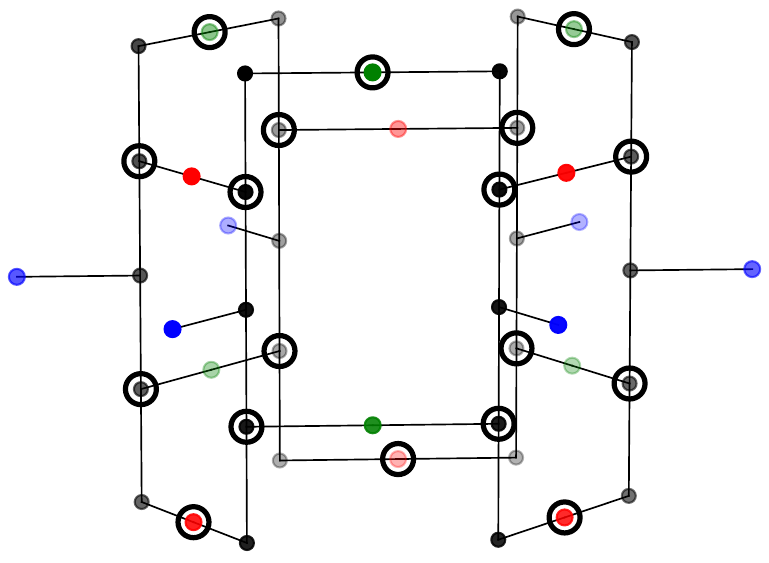}%
        }%
    }%
\end{minipage}%
\hfill
\begin{minipage}[t]{0.20\textwidth}
    \centering
    \stackinset{l}{-15pt}{t}{2pt}{\textbf{(b)}}{%
        \parbox[c][5.3cm][c]{\linewidth}{%
            \centering
            \includegraphics[width=0.95\linewidth]{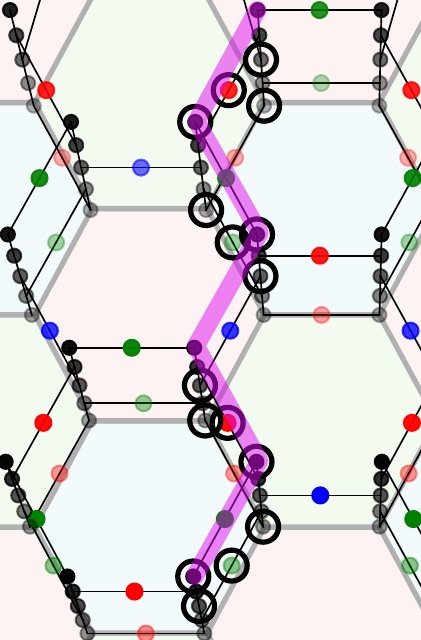}%
        }%
    }%
\end{minipage}%
\hfill
\begin{minipage}[t]{0.45\textwidth}
    \centering
    \stackinset{l}{-12pt}{t}{2pt}{\textbf{(c)}}{%
        \parbox[c][5.3cm][c]{\linewidth}{%
            \centering
            \includegraphics[width=0.95\linewidth]{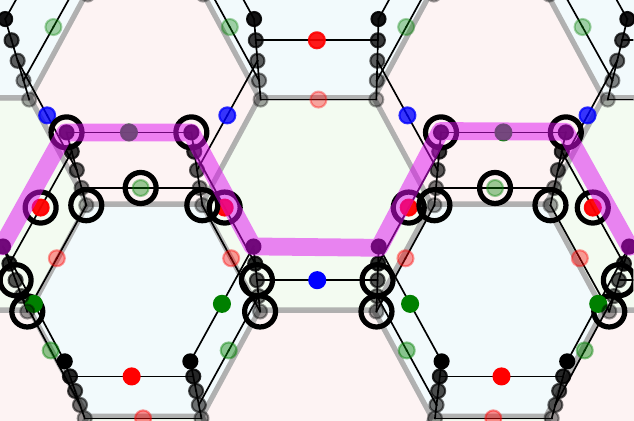}%
        }%
    }%
\end{minipage}
\caption{
(a) An 18-qubit \(X\)-type check of the FFCC code. Its support spans five consecutive layers in the sequence SL--data--no support--data--SL, with the supported qubits circled.
(b)--(c) Vertical and horizontal logical-\(X\) correlation surfaces, respectively. In each panel, the magenta line indicates the reference path used to construct the surface, and the circled qubits indicate its support. On odd-numbered layers (\(1,3,\ldots\)), the surface contains data qubits along the reference path that are connected to SL qubits on that path. On even-numbered layers (\(2,4,\ldots\)), it contains SL qubits along the reference path that are connected to two data qubits on that path. This construction requires an odd total number of layers.
}
\label{fig:ffcc_checks_logicals}
\end{figure*}

\section{The codes}

\label{section:code_construction}

\paragraph{RHG code:} \label{subsec:rhg} The RHG code, introduced by Raussendorf, Harrington, and Goyal, is implemented using a three-dimensional cluster-state resource for fault-tolerant MBQC~\cite{raussendorf2006aop,fowler2008arxiv}. 
A cluster state is constructed by initializing each qubit in the $\ket{+}$ state and applying CZ gates along the edges of the desired graph. This gives stabilizers of the form \mbox{$S_{i} = X_{i} \prod_{j \in N(i)} Z_{j}$}, where the product runs over the neighbors of qubit $i$~\cite{briegel2001persistent,hein2004multiparty}. 
The fundamental objects used for error correction in MBQC are \emph{checks} and \emph{correlation surfaces}, which are obtained from products of cluster-state stabilizers and correspond to deterministic products of single-qubit $X$-basis measurement outcomes~\cite{raussendorf2007topological,fowler2008arxiv}.

The RHG code can be viewed as a foliated version of the surface code, in which successive two-dimensional surface-code layers are stacked and coupled to form a three-dimensional cluster state~\cite{bolt2016prl}. With periodic boundary conditions, this construction becomes a
foliated toric code, which we analyze here.
Accordingly, the code can be described layer by layer. 
The CZ edges within each layer are called space-like edges, while those connecting adjacent layers are called time-like edges. Qubits that participate in both types of edges are called data qubits, whereas qubits connected only by space-like edges are called syndrome-like (SL) qubits. These names reflect their roles in the equivalent layered surface-code picture: data qubits carry the logical information, while SL qubits act as ancillas whose measurements provide syndrome information. 

Fig.~\ref{fig:rhg} shows the RHG code and a 6-qubit check. The logical-$X$ correlation surface lies along data qubits on odd-numbered layers ($1,3,\ldots$), as illustrated in panel (a). We choose periodic boundary conditions for RHG to place all three codes on the same footing and avoid boundary-dependent finite-size effects in the LER comparison. The required wraparound connections are compatible with photon-based routing in the compound architecture. The resulting toric construction encodes two logical qubits. By symmetry between the two periodic
spatial directions, it is sufficient to analyze one of them.

Although rotated-surface-code foliations can reduce the qubit count, we use the standard RHG construction to avoid imposing hook-error-sensitive gate-order constraints that could restrict more flexible generation strategies~\cite{tomita2014lowdistance,kishony2026offhook}.

\paragraph{FFCC code:} \label{subsec:ffcc} 

The FFCC code was introduced by Paesani and Brown~\cite{paesani2023prl}. Its name, Foliated Floquet Color Code, reflects its three-colorable lattice, as in the color code~\cite{bombin2006prl,bombin2007pra}, its inspiration from Floquet color codes~\cite{hastings2021quantum,davydova2023prx}, and its layered structure~\cite{bolt2016prl}. In Floquet color codes, a time-periodic measurement protocol dynamically generates logical qubits.

The FFCC code is realized as a three-dimensional cluster state built from multiple layers of a hexagonal lattice, as shown in Fig.~\ref{fig:ffcc_cluster_layers}(a). As in the RHG code, we assume periodic boundary conditions within each layer and distinguish data qubits from SL qubits according to their participation in time-like and space-like edges.

The underlying hexagonal lattice is three-colorable, with data qubits occupying its vertices. SL qubits are placed in successive layers according to the three colors of the lattice: red, green, and blue. In a layer corresponding to a given color, SL qubits are placed along the line connecting pairs of hexagonal faces of that color and connected to the data qubits at the endpoints of that line, as illustrated for the red layer in Fig.~\ref{fig:ffcc_cluster_layers}(b). This pattern repeats periodically~\cite{paesani2023prl}.

We impose periodic boundary conditions by identifying opposite sides of a rectangular cut through the hexagonal lattice. The cut is chosen to make its aspect ratio as close to unity as possible, as shown in Fig.~\ref{fig:ffcc_cluster_layers}(b). The lattice spans $4+2m$ hexagons along the horizontal direction and $3+3n$ along the vertical direction. Defining the linear size as $l$, we use dimensions $(l,l)$ for even $l$, corresponding to $3n-2m=1$. For odd $l$, we use $(l,l+1)$, corresponding to $3n-2m=2$.

The error-correction checks are 18-qubit $X$-type measurements supported across five consecutive layers, as illustrated in Fig.~\ref{fig:ffcc_checks_logicals}(a). Near the temporal boundaries, their support is reduced to 15 or 12 qubits. An additional check is oriented perpendicular to the temporal direction and is supported only on SL qubits. We do not use this check in our error-correction procedure. See Fig.~2(c) of Paesani and Brown~\cite{paesani2023prl}.

With periodic spatial boundary conditions, the FFCC encodes two logical qubits. 
A logical-$X$ correlation surface is constructed by following a horizontal or vertical path around the periodic lattice through all foliation layers. Its support alternates between data qubits on odd-numbered layers and SL qubits on even-numbered layers, as detailed in Figs.~\ref{fig:ffcc_checks_logicals}(b)--(c). We name each logical qubit according to the orientation of its logical-$X$ surfaces. The horizontal logical has a horizontal logical-$X$ and a vertical logical-$Z$, while the vertical logical has a vertical logical-$X$ and a horizontal logical-$Z$. The corresponding logical-$Z$ construction uses a layer pattern shifted by one foliation layer.

\paragraph{Reduced FFCC code:} \label{subsec:reduced_ffcc} 

\begin{figure}[!t]
\centering
\stackinset{l}{0pt}{t}{2pt}{(a)}{%
    \makebox[\linewidth][c]{%
        \includegraphics[width=0.8\linewidth]{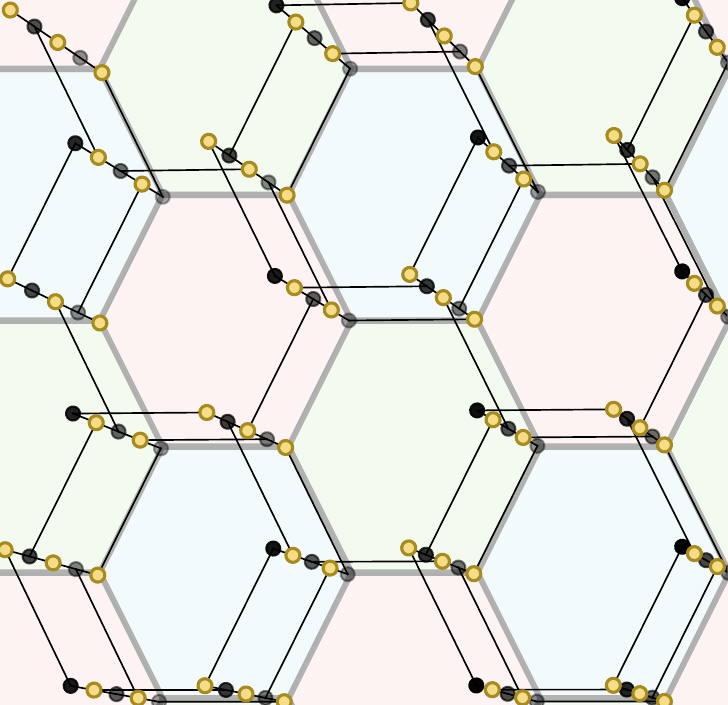}%
    }%
}
\par\medskip
\stackinset{l}{0pt}{t}{2pt}{(b)}{%
    \makebox[\linewidth][c]{%
        \includegraphics[width=0.55\linewidth]{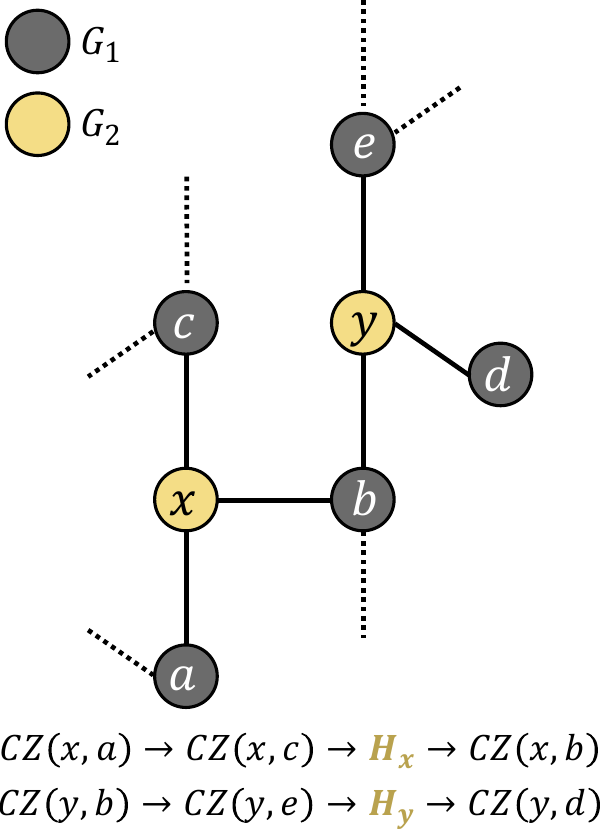}%
    }%
}
\caption{
(a) Reduced FFCC code construction for $l=3$ and $t=5$, with periodic boundary conditions imposed on a rectangular cut. The graph is bipartitioned into the two sets \(G_1\) (gray) and \(G_2\) (yellow). There are no SL qubits. Edges denote CZ gates; periodic-boundary edges are omitted for clarity. (b) Local subgraph containing two $G_2$ qubits, $x$ and $y$, that are next-nearest neighbors, together with the adjacent $G_1$ qubits $a,b,c,d$, and $e$. Each $G_2$ qubit undergoes an $H$ gate between its interlayer and intralayer CZ gates.
}
\label{fig:reduced_ffcc_structure}
\end{figure}

\begin{figure*}
\centering
\begin{minipage}[t]{0.3\textwidth}
    \centering
    \stackinset{l}{2pt}{t}{2pt}{\textbf{(a)}}{%
        \parbox[c][4.6cm][c]{\linewidth}{%
            \centering
            \includegraphics[width=0.95\linewidth]{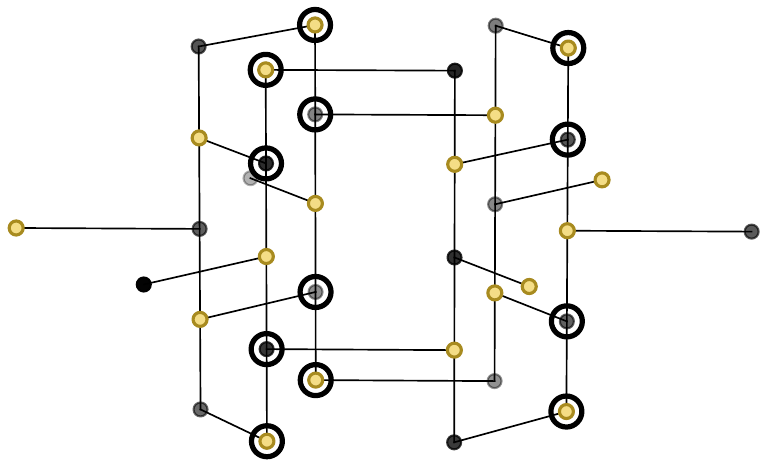}%
        }%
    }%
\end{minipage}%
\hfill
\begin{minipage}[t]{0.20\textwidth}
    \centering
    \stackinset{l}{-15pt}{t}{2pt}{\textbf{(b)}}{%
        \parbox[c][4.6cm][c]{\linewidth}{%
            \centering
            \includegraphics[width=0.95\linewidth]{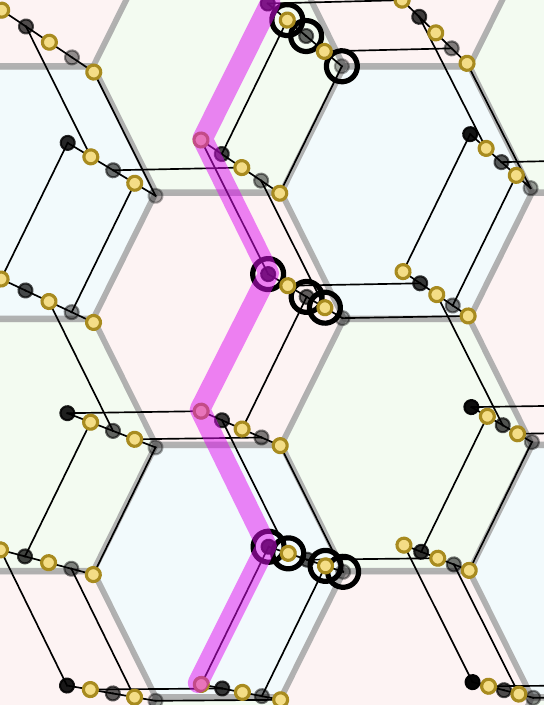}%
        }%
    }%
\end{minipage}%
\hfill
\begin{minipage}[t]{0.45\textwidth}
    \centering
    \stackinset{l}{-12pt}{t}{2pt}{\textbf{(c)}}{%
        \parbox[c][4.6cm][c]{\linewidth}{%
            \centering
            \includegraphics[width=0.95\linewidth]{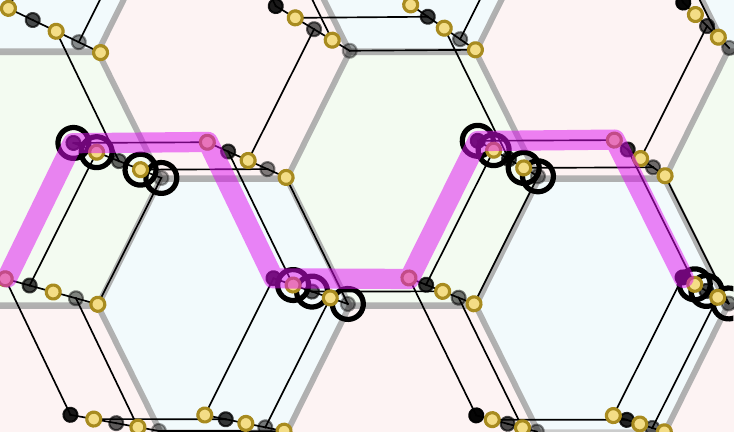}%
        }%
    }%
\end{minipage}

\caption{
(a) A 12-qubit check of the reduced FFCC code. Its support spans five consecutive layers in the sequence \(G_2\)--\(G_1\)--no support--\(G_1\)--\(G_2\), with the supported qubits circled.
(b)--(c) Vertical and horizontal logical-\(X\) correlation surfaces, respectively. In each panel, the magenta line indicates the reference path used to construct the surface, and the circled qubits indicate its support. The surface contains qubits along paths parallel to the reference path: \(G_1\) qubits on odd-numbered layers (\(1,3,\ldots\)) and \(G_2\) qubits on even-numbered layers (\(2,4,\ldots\)). Only qubits connected to opposite-partition qubits along the same path are included. This construction requires an odd total number of layers.
}
\label{fig:reduced_ffcc_checks_logicals}
\end{figure*}

The reduced FFCC code~\cite{paesani2023prl} is obtained by replacing each FFCC SL qubit with an $H$ gate on one of its neighboring data qubits. The resulting construction is no longer a standard cluster-state preparation, and its checks and logical operators are smaller than those of FFCC, as illustrated in Fig.~\ref{fig:reduced_ffcc_checks_logicals}. With periodic boundary conditions, reduced FFCC also encodes two logical qubits, with logical operators defined analogously to those of FFCC.
 
There are multiple possible $H$ gate assignments. We use the one best suited to our compound architecture, where $H$ gates must be applied to photons rather than atoms. The corresponding local gate ordering is shown in Fig.~\ref{fig:reduced_ffcc_structure}(b), and the hardware constraints are discussed in Sec.~\ref{section:generation_schemes}.

\section{Generation schemes}
\label{section:generation_schemes}

\begin{figure*}[!t]
    \centering

    \stackinset{l}{-20pt}{t}{0pt}{\textbf{(a)}}{%
        \includegraphics[width=0.85\textwidth]
        {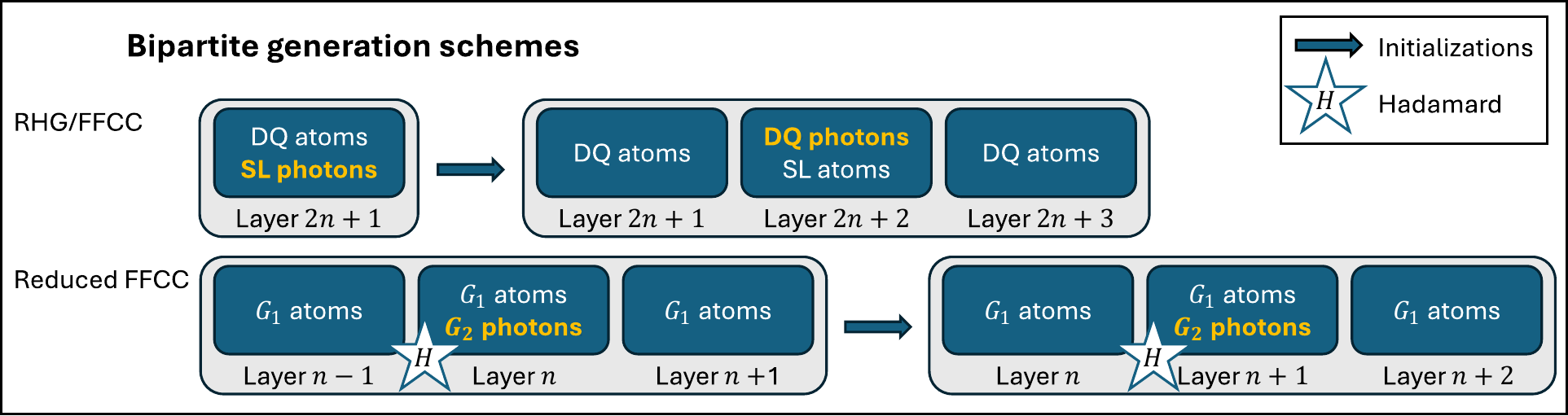}%
    }

    \vspace{0.8em}

    \stackinset{l}{-20pt}{t}{0pt}{\textbf{(b)}}{%
        \includegraphics[width=0.85\textwidth]
        {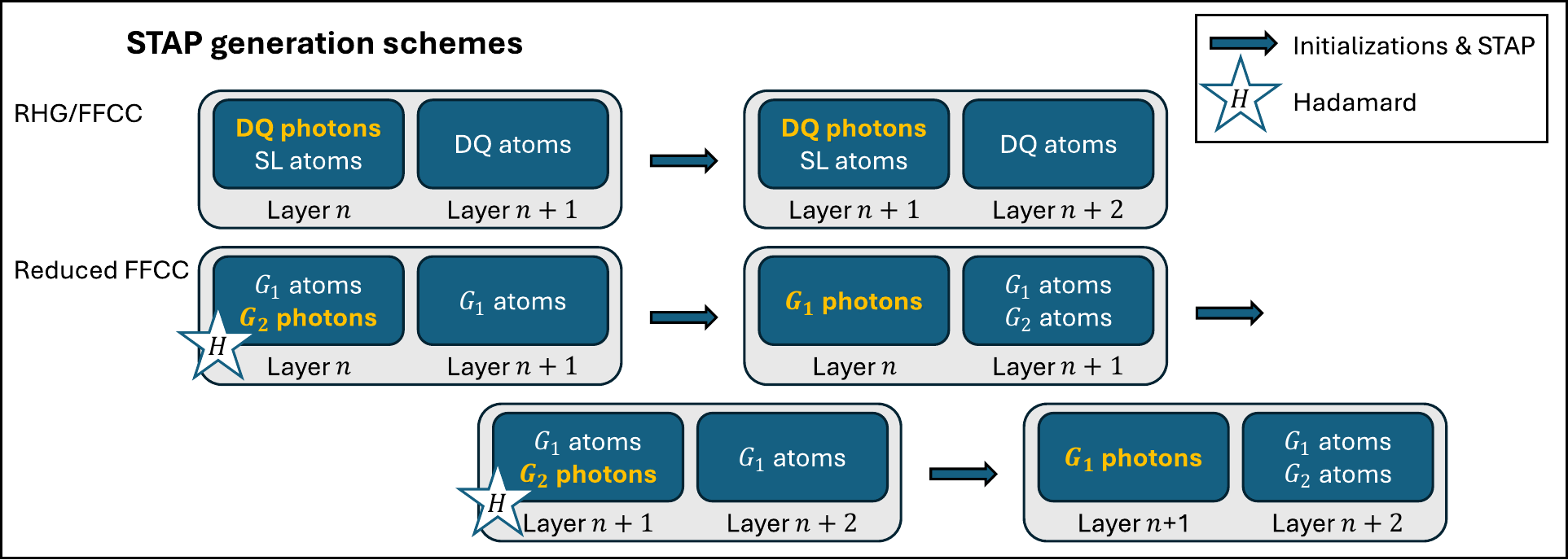}%
    }

    \caption{
    Generation schemes for the RHG, FFCC, and reduced FFCC codes.
    (a) Bipartite scheme. For each code, two consecutive stages are shown. Photons assigned to a stage complete their cycles before the schedule advances to the next stage. The arrow marks the interval during which the atoms required for the next stage are initialized. In RHG and FFCC, the photonic roles alternate between SL-qubit and data-qubit (DQ) positions in consecutive layers. In reduced FFCC, \(G_2\) is the photonic partition in the stages shown. Because each atom can participate in only one CZ gate per cycle, the number of photon cycles required at each stage is set by the maximum number of participating photons coupled to the same atom. RHG requires two and four cycles for the stages of layers \(2n+1\) and \(2n+2\), respectively. FFCC requires one and two cycles, while reduced FFCC requires one cycle per layer.
    (b) STAP scheme. The arrows mark the atom-initialization and STAP operations that begin the next stage. In RHG and FFCC, atomic data qubits in layer \(n\) undergo STAP and complete their photon cycles, while the SL qubits in layer \(n\) and the data qubits in layer \(n+1\) remain atomic. Each stage requires four cycles for RHG and two for FFCC. In reduced FFCC, each layer is processed in two one-cycle stages. The \(G_2\) qubits first undergo STAP and complete a cycle that includes the photonic \(H\) gate. The \(G_1\) qubits then undergo STAP and complete their cycle.
    }
    \label{fig:generation_schemes}
\end{figure*}

A \textit{generation scheme} specifies how the target code graph or circuit is assembled from the native hardware operations. These operations are atom initialization, photon generation, photon--atom CZ gates, photon measurement, STAP, and photonic single-qubit gates~\cite{qs2026arxiv,aqua2025}. STAP transfers the quantum state of an atom to an emitted photon. Atomic measurement is implemented by STAP followed by photon detection. In our circuit-level loss model, photon generation, photon participation in a CZ gate, STAP, and photon measurement are each assigned loss probability \(p\). Atomic measurement is assigned \(2p\), approximating the two sequential lossy steps, STAP and photon detection. Atom initialization is treated as loss-free because failure is heralded and corrected before any interaction. Photonic \(H\) gates are also treated as loss-free, with optical-path loss assigned to the subsequent lossy operation. We focus on photon loss as the dominant error mechanism. Other error channels and the rationale for neglecting them are discussed in the Supplementary
Material. The model is summarized in Table~\ref{tab:error_model}.

The hardware supports CZ gates only between photons and atoms. Photons complete their CZ sequence in a single pass, whereas atoms can participate in only one CZ gate at a time. We therefore describe the schedules from the photon perspective in terms of \textit{photon cycles}. A photon cycle begins with photon initialization or STAP, includes the required CZ gates and any photonic H gate, and ends with photon measurement~\cite{qs2026arxiv}. We consider two generation schemes, the \textit{bipartite scheme} and the \textit{STAP scheme}, shown in Fig.~\ref{fig:generation_schemes}(a) and (b), respectively.

\paragraph{Bipartite scheme:} As all three codes are bipartite, we can assign one graph partition to photons and the other to atoms. Each qubit is initialized in a $\ket{+}$ state immediately before its first CZ gate and measured in the X-basis immediately after its last CZ gate. The schedule advances approximately layer by layer through successive photon cycles.

For RHG and FFCC, the photon and atom assignments alternate between SL-qubit and data-qubit positions in consecutive layers. The photons then perform the required CZ gates layer by layer. For reduced FFCC, which has no SL qubits, the local ordering of CZ and H gates is essential, as shown in Fig.~\ref{fig:reduced_ffcc_structure}(b). In each layer, photons perform CZ gates first with the previous layer and then with the next, undergo an H gate, and finally perform the intralayer CZ gates. 

In the bipartite scheme, the maximum number of simultaneously active atoms for RHG and FFCC equals one full layer plus the data qubits of another layer. These atoms span at most three layers. For reduced FFCC, the maximum number of simultaneously active atoms equals the qubit count of one and a half layers, again spanning at most three layers.

\paragraph{STAP scheme:} In the STAP scheme, all qubits are initialized as atoms and measured as photons, and STAP is applied to every qubit. The CZ schedule is arranged approximately layer by layer, and qubits are measured once their required CZ gates are complete. 

For the RHG and FFCC codes, data qubits in each bulk layer are initialized as atoms, perform one CZ gate with a photonic data qubit from the previous layer, undergo STAP, and then perform their remaining CZ gates as photons. The SL qubits perform all their CZ gates as atoms and undergo STAP only before measurement.
For reduced FFCC, the code is constructed layer by layer, as illustrated in Fig.~\ref{fig:reduced_ffcc_structure}(a), which shows five layers arranged bottom to top. $G_1$ and $G_2$ denote the two graph partitions, as shown in Fig.~\ref{fig:reduced_ffcc_structure}(b). In the first layer, the $G_2$ qubits undergo STAP, perform CZ gates with the second layer, undergo an $H$ gate, complete the intralayer CZ gates in the first layer, and are measured. The $G_1$ qubits then undergo STAP, perform the interlayer CZ gates with the second layer, and are measured.

For the STAP scheme, the maximum number of simultaneously active atoms is the same as in the bipartite scheme, but these atoms span at most two layers.

\section{Results and discussion}
\label{section:results}

\begin{table}
\centering
\begin{tabular}{lcc}
\hline
\textbf{Operation} & \textbf{Photon} & \textbf{Atom} \\
\hline
Initialization           & $p$   & $0$  \\
CZ (photon--atom)        & $p$   & $0$  \\
STAP                   & --    & $p$  \\
Measurement              & $p$   & $2p$ \\
\hline
\end{tabular}
\caption{Circuit-level loss model used in the threshold simulations~\cite{qs2026arxiv}. We focus solely on photon loss as the dominant error mechanism in our hardware. For simplicity, all lossy operations are modeled with a single photon-loss parameter $p$, except atomic measurement, which includes both STAP and photon detection.}
\label{tab:error_model}
\end{table}

\begin{table}
\centering
\begin{tabular}{lccc}
\hline
\textbf{Code} & \textbf{IID} & \multicolumn{2}{c}{\textbf{Circuit-level}} \\
\cline{3-4}
&  & \textbf{Bipartite} & \textbf{STAP} \\
\hline
RHG           & 0.25  & 0.0275 & 0.0275 \\
FFCC          & 0.085 & 0.0145 & 0.0125 \\
Reduced FFCC  & 0.135 & 0.0185 & 0.015 \\
\hline
\end{tabular}
\caption{IID and circuit-level loss thresholds under the error model of Table~\ref{tab:error_model}. The IID thresholds are taken from previous studies~\cite{Barrett_2010,paesani2023prl}. The circuit-level values are obtained from logical-\(X\) memory simulations shown in Figs.~\ref{fig:thresholds_rhg}, \ref{fig:thresholds_ffcc}, and \ref{fig:thresholds_reduced_ffcc}. In the bipartite scheme, we use an atom-supported correlation surface. For FFCC and reduced FFCC, the listed circuit-level thresholds correspond to even \(l\).}
\label{tab:thresholds_summary}
\end{table}

\begin{figure}[!t]
\centering

\stackinset{l}{-5pt}{t}{2pt}{\textbf{(a)}}{%
    \includegraphics[width=0.9\linewidth]{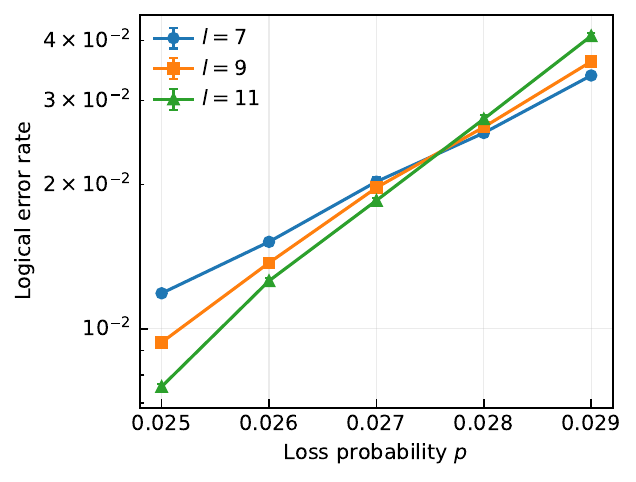}%
}%

\vspace{0.8em}

\stackinset{l}{-5pt}{t}{2pt}{\textbf{(b)}}{%
    \includegraphics[width=0.9\linewidth]{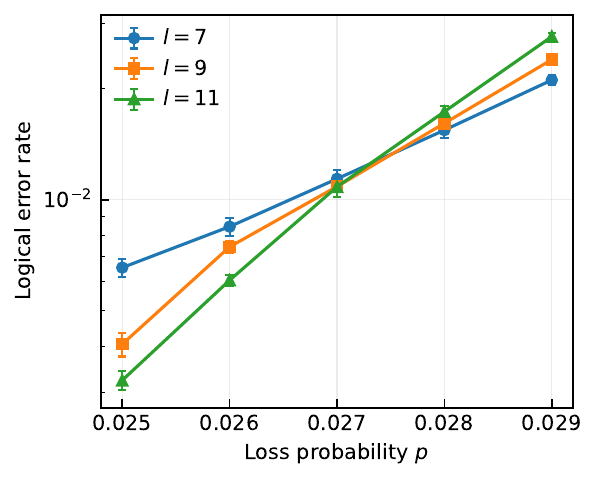}%
}%

\caption{Circuit-level logical-\(X\) memory threshold simulations for the RHG code, shown for the bipartite (a) and STAP (b) generation schemes.
Each data point uses \(3{,}000\) loss realizations and \(10{,}000\) decoding iterations per realization.
Error bars denote the standard error of the mean (SEM) from grouped loss-realization batches.}
\label{fig:thresholds_rhg}
\end{figure}

\begin{figure*}[!t]
\centering

\begin{minipage}[t]{0.46\textwidth}
    \centering
    \vspace{0pt}
    \stackinset{l}{0pt}{t}{2pt}{\textbf{(a)}}{%
        \includegraphics[width=\linewidth]{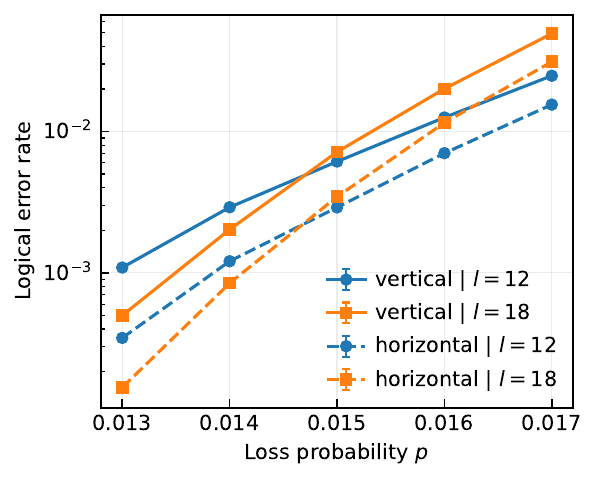}%
    }%
\end{minipage}
\hfill
\begin{minipage}[t]{0.46\textwidth}
    \centering
    \vspace{0pt}
    \stackinset{l}{0pt}{t}{2pt}{\textbf{(b)}}{%
        \includegraphics[width=\linewidth]{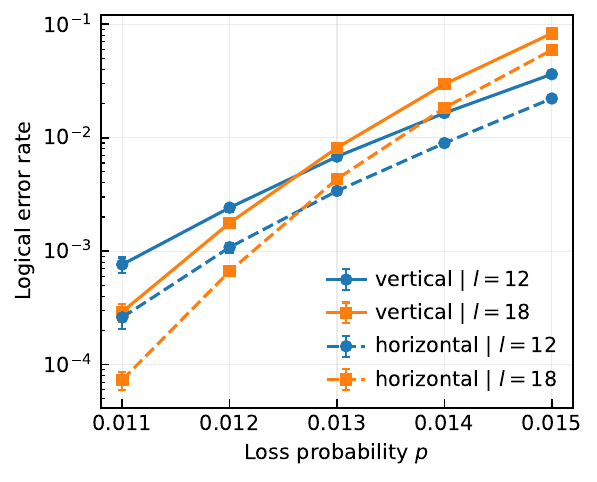}%
    }%
\end{minipage}

\caption{Circuit-level logical-\(X\) memory threshold simulations for the FFCC code with even $l$. 
Panel (a) shows the bipartite generation scheme, and panel (b) shows the STAP generation scheme. 
In both panels, the solid line corresponds to the vertical logical operator and the dashed line corresponds to the horizontal logical operator. 
The two logical orientations give the same threshold within numerical precision. Each point uses \(3{,}000\) loss realizations and \(10{,}000\) decoding iterations per realization. Error bars show SEM. Odd-$l$ results are given in Fig.~\ref{fig:thresholds_ffcc_odd}.}
\label{fig:thresholds_ffcc}
\end{figure*}

\begin{figure*}[!t]
\centering

\begin{minipage}[t]{0.46\textwidth}
    \centering
    \vspace{0pt}
    \stackinset{l}{0pt}{t}{2pt}{\textbf{(a)}}{%
        \includegraphics[width=\linewidth]{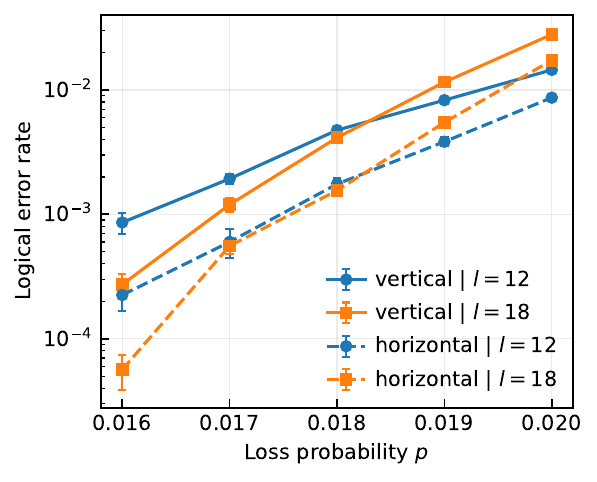}%
    }%
\end{minipage}
\hfill
\begin{minipage}[t]{0.46\textwidth}
    \centering
    \vspace{0pt}
    \stackinset{l}{0pt}{t}{2pt}{\textbf{(b)}}{%
        \includegraphics[width=\linewidth]{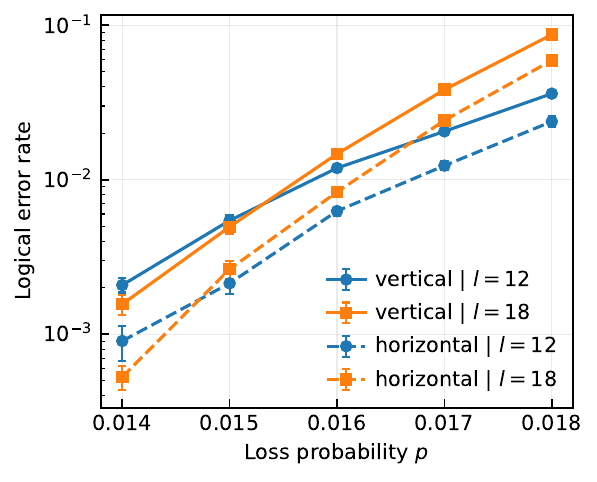}%
    }%
\end{minipage}

\caption{Circuit-level logical-\(X\) memory threshold simulations for the reduced FFCC code with even $l$. 
Panel (a) shows the bipartite generation scheme, and panel (b) shows the STAP generation scheme. 
In both panels, the solid line corresponds to the vertical logical operator and the dashed line corresponds to the horizontal logical operator. 
The two orientations give the same threshold within numerical precision. 
Each point uses \(3{,}000\) loss realizations and \(10{,}000\) decoding iterations per realization.
Error bars show SEM. Odd-$l$ results are given in Fig.~\ref{fig:thresholds_reduced_ffcc_odd}.}
\label{fig:thresholds_reduced_ffcc}
\end{figure*}

\paragraph{Thresholds:} \label{par:thresholds} We first compare the thresholds of the three codes under IID loss and under the circuit-level loss model of Table~\ref{tab:error_model}. Table~\ref{tab:thresholds_summary} summarizes the resulting thresholds for all three codes and both generation schemes. For FFCC and reduced FFCC, the parity of \(l\) slightly affects the threshold because it changes the layer shape: even \(l\) gives a square geometry, whereas odd \(l\) gives a rectangular one. Since the even-\(l\) thresholds are higher, the table uses the even-\(l\) case for comparison with RHG. The odd-\(l\) results are given in the Supplementary Material, Figs.~\ref{fig:thresholds_ffcc_odd} and~\ref{fig:thresholds_reduced_ffcc_odd}.

The baseline circuit-level thresholds in
Table~\ref{tab:thresholds_summary} and
Figs.~\ref{fig:thresholds_rhg}--\ref{fig:thresholds_reduced_ffcc}
are obtained from single logical-\(X\) memory simulations.
In the bipartite scheme, we choose the photon--atom assignment so that
the simulated logical-\(X\) correlation surface is supported on atoms.
This conservative choice is motivated by the lower atom-supported RHG
threshold found previously~\cite{qs2026arxiv}, where photon-loss-induced bond loss is more detrimental than in the photon-supported case.
In the STAP scheme, qubits change physical type during the schedule, so
there is no analogous fixed atom-supported or photon-supported
correlation surface.

Under circuit-level loss, the IID loss budget is distributed across the physical operations acting on each qubit. Mid-circuit photon loss can also remove subsequent bonds and induce errors on neighboring atoms. The resulting threshold therefore depends on both graph degree, which determines the number of CZ gates per qubit, and the generation scheme, which determines their ordering. RHG shows the largest reduction from the IID to the circuit-level threshold (from $25\%$ to $2.75\%$), consistent with its largest graph degree, while FFCC shows the smallest reduction, consistent with its smallest graph degree.

Among the codes considered here, RHG nevertheless achieves the highest threshold of \(\sim2.75\%\). Scheme choice has a negligible effect on the RHG threshold, a modest effect on the FFCC and reduced FFCC thresholds, and a clearer effect on the LER. Because every data qubit in the STAP scheme performs one CZ gate while atomic, this scheme reduces exposure to lossy photon--atom CZ gates and may reduce loss-time ambiguity for the decoder, without adding a net non-CZ loss opportunity. Conversely, because STAP occurs before the CZ sequence is complete, loss during STAP can remove subsequent bonds and affect neighboring atoms. These competing effects make the net change code dependent. The LER comparison can also depend on the logical state considered, since the bipartite scheme can assign complementary correlation surfaces to different physical qubit types, whereas STAP does not. For RHG, this effect was demonstrated numerically in
previous work~\cite{qs2026arxiv}. Further
details are given in the Supplementary Material.

\begin{figure}
\centering
\includegraphics[width=0.95\linewidth]
{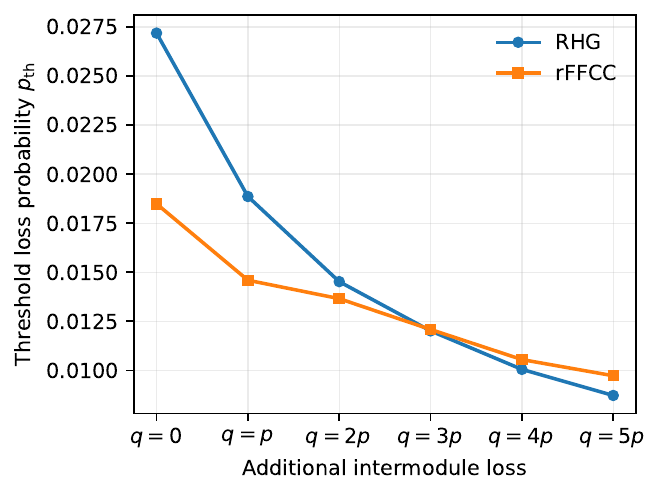}
\caption{
Effect of additional loss on intermodule CZ connections.
The simulations use the circuit-level loss model of
Table~\ref{tab:error_model}, with CZ connections designated as intermodule having a loss probability
\(p_{\mathrm{inter}}=p+q\) instead of \(p\). This modification is applied to the third and fourth CZ connections in RHG and only to the third CZ connection in reduced FFCC. The thresholds $p_{\mathrm{th}}$ are shown for $q\in\{0,p,\ldots,5p\}$ and are extracted from the finite-distance crossings shown in Supplementary Fig.~\ref{fig:intermodule_loss_ler}.
}
\label{fig:intermodule_loss_comparison}
\end{figure}

\paragraph{Intermodule loss:} \label{par:intermodule_loss} 
The comparison above assumes a uniform loss probability for all photon--atom CZ gates. This need not hold in a modular implementation, where intermodule connections may traverse additional couplers, switches, or delay lines and therefore incur extra optical loss~\cite{bartolucci2021switchnetworksphotonicfusionbased,bombin2021interleaving,aghaeerad2025scaling}.
We model this asymmetry by assigning loss probability \(p\) to
intramodule CZ gates and \(p_{\mathrm{inter}}=p+q\) to intermodule CZ gates, where \(q\geq0\) is the excess intermodule loss.

Several photonic constructions use locally generated resource states that are subsequently connected by inter-resource operations. In these constructions, each bulk qubit commonly has two graph connections
within its local resource, with the remaining connections linking different resources. In previous FFCC constructions~\cite{paesani2023prl,chan2025prxquantum},
single emitters generate linear cluster states that are then connected
by fusion measurements. A related surface-code construction instead uses six-qubit ring resource states connected by inter-resource fusions~\cite{bartolucci2023fusion}. The circled qubits in Fig.~\ref{fig:rhg} similarly indicate a possible partition of the RHG lattice into six-ring unit cells. Motivated by this two-internal-link structure, we model the first two CZ gates of each qubit as intramodule and the remaining CZ gates as the more lossy intermodule connections.

As shown in Fig.~\ref{fig:intermodule_loss_comparison}, both thresholds decrease with increasing \(q\), but RHG degrades more rapidly because two CZ connections per photon carry the intermodule loss penalty, compared with one in reduced FFCC. The ordering reverses between the sampled cases \(q=2p\) and \(q=3p\): RHG has the higher threshold at lower intermodule loss, whereas reduced FFCC has the higher threshold at larger \(q\).

This result complements previous studies in which low-degree Floquet codes benefited from native two-qubit measurements, probabilistic fusion, or photonic repeat-until-success
operations~\cite{gidney2021honeycomb,
dessertaine2026enhancedfaulttolerancephotonicquantum,
paesani2023prl,chan2025prxquantum}. Here, by contrast, the native
photon--atom CZ primitive is held fixed, and the code ordering reverses when additional loss is assigned to selected intermodule CZ connections.

For small loss probabilities, \(q\approx2p\) can arise from two
additional loss channels of order \(p\), such as out-coupling and
in-coupling at a module boundary. Larger values are physically
plausible when the intermodule path also contains active switches,
connectors, or delay-line propagation
~\cite{bartolucci2021switchnetworksphotonicfusionbased,
bombin2021interleaving,aghaeerad2025scaling}. In this regime, the
smaller number of intermodule CZ connections in reduced FFCC can
outweigh its lower uniform-loss threshold.


\begin{figure}[!t]
\centering

\begin{minipage}{\linewidth}
    \centering
    \stackinset{l}{0pt}{t}{2pt}{\textbf{(a)}}{%
        \includegraphics[width=0.9\linewidth]{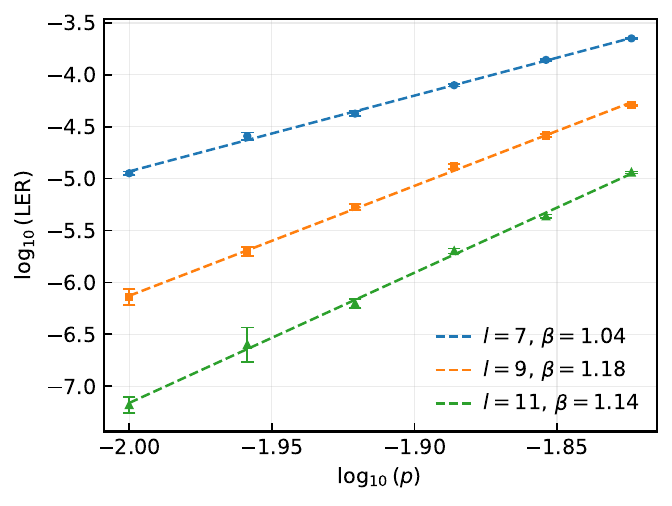}%
    }%
\end{minipage}

\vspace{0.8em}

\begin{minipage}{\linewidth}
    \centering
    \stackinset{l}{0pt}{t}{2pt}{\textbf{(b)}}{%
        \includegraphics[width=0.9\linewidth]{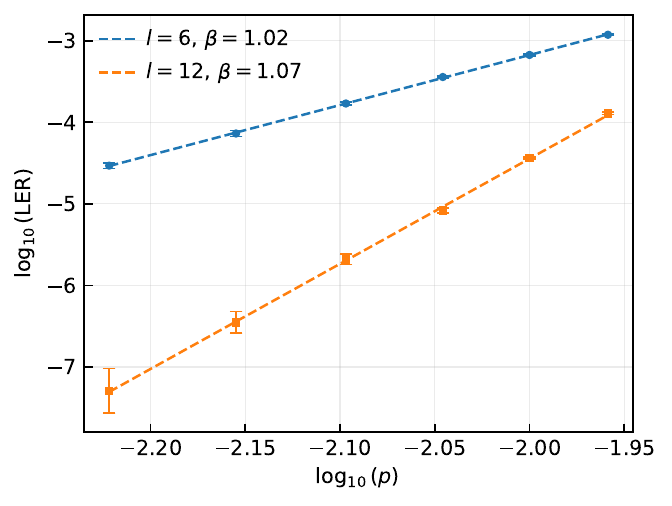}%
    }%
\end{minipage}

\vspace{0.8em}

\begin{minipage}{\linewidth}
    \centering
    \stackinset{l}{0pt}{t}{2pt}{\textbf{(c)}}{%
        \includegraphics[width=0.9\linewidth]{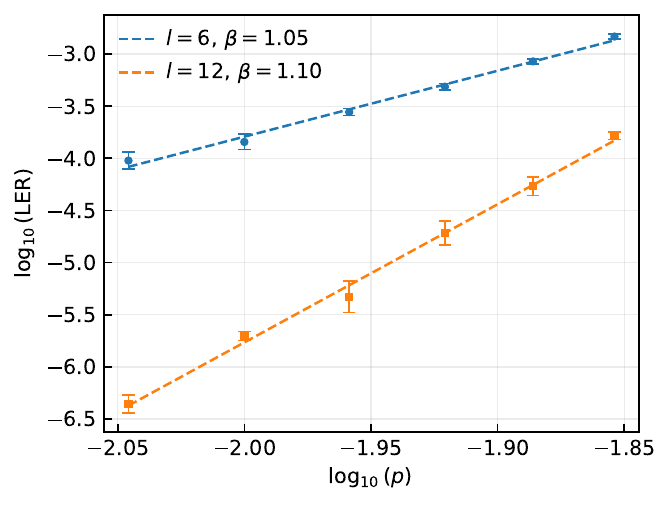}%
    }%
\end{minipage}

\caption{LER scaling for the RHG, FFCC, and reduced FFCC codes in the bipartite scheme. Fits give \(\beta\) slightly above 1 for all three codes. Each point uses at least \(50{,}000\) loss realizations and \(10{,}000\) decoding iterations per realization. Error bars show SEM.}
\label{fig:scaling_bipartite_all_codes}
\end{figure}

\paragraph{LER scaling:} \label{par:scaling} 
We next study how the LER scales with distance in the bipartite
scheme. The code distance is the minimum physical-qubit weight of a
nontrivial logical operator. For our circuit-level detector model, we estimate the corresponding graphlike fault distance using
\texttt{shortest\_graphlike\_error()} in 
Stim~\cite{gidney2021stim}. We obtain \(d=l\) for RHG, and for FFCC and reduced FFCC, \(d=l\) for even \(l\) and \(d=l+1\) for odd \(l\). We use these graphlike-distance estimates as \(d\) in the scaling analysis below.

Because \texttt{shortest\_graphlike\_error()} omits ungraphlike
detector-error mechanisms by default, its result is an upper bound on the true circuit fault distance. We therefore characterize the
near-threshold LER scaling as an independent consistency check on the graphlike-distance assignments. We model the LER as
\begin{equation}
\mathrm{LER}(p,d)
=
\mathrm{LER}(p_{\mathrm{th}})
\left(
    \frac{p}{p_{\mathrm{th}}}
\right)^{\beta d},
\label{eq:ler_scaling}
\end{equation}
where $p_{\mathrm{th}}$ is the threshold and $\beta$ is the scaling exponent~\cite{watson2014njp,lee2025quantum}. 
Equivalently,
\begin{equation*}
\ln\!\left(\mathrm{LER}(p,d)\right)
=
\ln\!\left(\mathrm{LER}(p_{\mathrm{th}})\right)
+
\beta d \ln\!\left(p/p_{\mathrm{th}}\right).
\end{equation*}
For fixed $d$, the slope of $\ln(\mathrm{LER})$ versus $\ln(p/p_{\mathrm{th}})$ is \(\beta d\). We therefore obtain
\(\beta\) by dividing the fitted slope by \(d\). 
For all three codes, the fitted \(\beta\) is slightly above \(1\), as shown in Fig.~\ref{fig:scaling_bipartite_all_codes}. 

The slightly elevated fitted \(\beta\) may reflect a preasymptotic
waterfall regime, in which the greater multiplicity of higher-weight failure mechanisms outweighs their additional power-law
suppression~\cite{gu2604scalable}. At sufficiently low \(p\), the
lowest-weight failures are expected to dominate. Under our
normalization, this would give \(\beta=1\) if the graphlike distance equals the true circuit fault distance. Directly sampling this regime is challenging because the corresponding LERs are extremely small. However, the fitted values \(\beta>1\) strongly support the graphlike-distance assignments: over the sampled range, we find no indication of the reduced effective distance that would result from dominant lower-weight ungraphlike logical mechanisms.

\begin{figure*}
\centering

\begin{minipage}[t]{0.46\textwidth}
\centering
\vspace{0pt}
\stackinset{l}{0pt}{t}{2pt}{\textbf{(a)}}{%
\includegraphics[width=0.99\linewidth]
{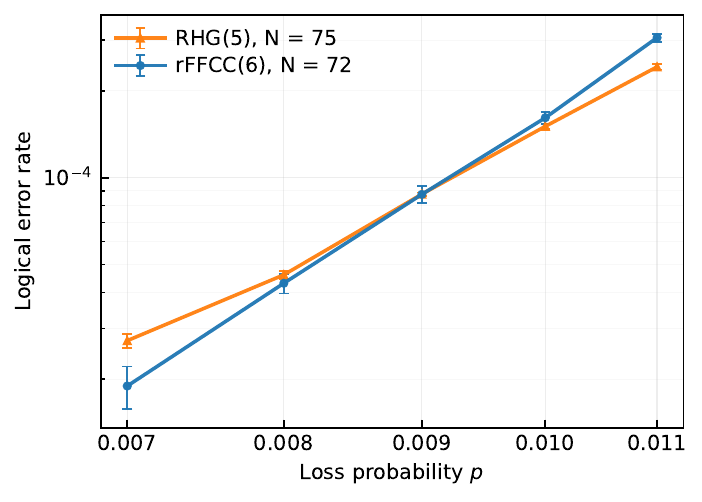}%
}%
\end{minipage}
\hfill
\begin{minipage}[t]{0.46\textwidth}
\centering
\vspace{0pt}
\stackinset{l}{-5pt}{t}{2pt}{\textbf{(b)}}{%
\includegraphics[width=0.96\linewidth]
{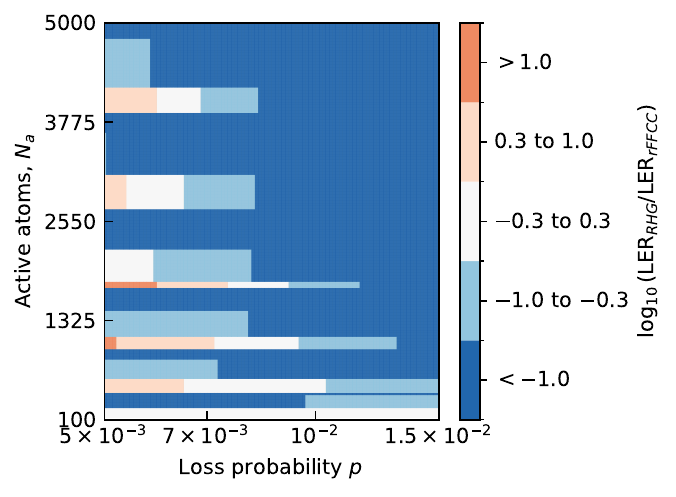}%
}%
\end{minipage}

\caption{
Comparison of code performance under matched resource budgets.
(a) Direct simulation at nearly matched physical-qubit counts per
layer: RHG\((d=5)\), \(N=75\), and reduced FFCC\((d=6)\), \(N=72\).
Error bars show SEM.
(b) Scaling-model comparison at fixed active-atom budget \(N_a\) over
\(100\leq N_a\leq5000\) and \(0.005\leq p\leq0.015\), showing
\(\log_{10}(\mathrm{LER}_{\mathrm{RHG}}/
\mathrm{LER}_{\mathrm{rFFCC}})\).
}
\label{fig:comparing_codes_subfigures}
\end{figure*}

\paragraph{Fixed resource budget:} \label{par:best_code_fixed_qubits} We finally compare below-threshold LERs under matched resource budgets.
We use direct numerical simulations at small distances and
scaling-model estimates based on Eq.~\eqref{eq:ler_scaling} for larger
systems. For simplicity, we restrict FFCC and reduced FFCC to even
\(l\), so all three code families satisfy \(d=l\). The number of
physical qubits per layer is
\begin{gather*}
N_{\mathrm{RHG}}(d) = 3d^2,\\
N_{\mathrm{FFCC}}(d) = 3d^2,\\
N_{\mathrm{rFFCC}}(d) = 2d^2.
\end{gather*}
At fixed distance, reduced FFCC requires fewer qubits than FFCC and
also has the higher circuit-level threshold. We therefore focus the
remaining comparison on RHG and reduced FFCC.

We first compare the two codes at nearly matched per-layer qubit
counts:
\begin{equation*}
N_{\mathrm{RHG}}(5)=75,
\qquad
N_{\mathrm{rFFCC}}(6)=72.
\end{equation*}
As shown in Fig.~\ref{fig:comparing_codes_subfigures}(a), reduced
FFCC\((6)\) has the lower LER for \(p\leq0.008\), the two codes are
comparable at \(p=0.009\), and RHG\((5)\) has the lower LER for
\(p\geq0.010\). Thus, even at nearly equal per-layer qubit count, the preferred code depends on the operating loss probability.

The per-layer qubit count, however, does not capture the atomic
resource requirement of the generation schedule. In the bipartite
scheme, the maximum numbers of simultaneously active atoms are
\begin{gather*}
N_{a,\mathrm{RHG}}(d) = 5d^2,\\
N_{a,\mathrm{rFFCC}}(d) = 3d^2,
\end{gather*}
as follows from the schedules in
Fig.~\ref{fig:generation_schemes}. To examine how this
hardware-specific overhead affects the comparison, we use
Eq.~\eqref{eq:ler_scaling} with \(\beta=1\) as a simplifying
approximation and extrapolate over
\begin{equation*}
100\leq N_a\leq5000,
\qquad
0.005\leq p\leq0.015.
\end{equation*}
The upper loss limit lies slightly below the reduced FFCC threshold, while the lower limit is chosen to keep the extrapolation reasonably close to the loss range sampled in the scaling simulations. For each code, we use the fitted threshold \(p_{\mathrm{th}}\) and an approximate value of \(\mathrm{LER}(p_{\mathrm{th}})\) obtained from
the threshold curves.

Specifically, for RHG we use
\begin{gather*}
p_{\mathrm{th}} = 0.0275,\\
\mathrm{LER}(p_{\mathrm{th}}) = 2.3\times10^{-2},\\
d = 5+2k,
\end{gather*}
and for reduced FFCC,
\begin{gather*}
p_{\mathrm{th}} = 0.0185,\\
\mathrm{LER}(p_{\mathrm{th}}) = 7.5\times10^{-3},\\
d = 6+6k,
\end{gather*}
where \(k=0,1,2,\ldots\). For RHG, the extrapolation is restricted to
the odd-distance family \(d\geq5\) used in our numerical simulations.
For each active-atom budget \(N_a\), we use the largest allowed
distance satisfying the corresponding resource constraint.

Figure~\ref{fig:comparing_codes_subfigures}(b) shows
\begin{equation*}
\log_{10}
\left(
\frac{\mathrm{LER}_{\mathrm{RHG}}}
     {\mathrm{LER}_{\mathrm{rFFCC}}}
\right).
\end{equation*}
Negative values indicate that RHG has the lower LER, while positive
values favor reduced FFCC. RHG performs better over most of the
parameter range considered, often by more than an order of magnitude. Reduced FFCC becomes favorable at sufficiently low \(p\) within specific active-atom-budget windows. The banded structure arises from the discrete allowed-distance families: reduced FFCC advances in steps of six, \(d=6+6k\), whereas RHG advances in steps of two, \(d=5+2k\). Immediately after a permitted increase in the reduced FFCC distance, its lower active-atom cost per unit distance can outweigh its lower loss tolerance over a finite resource window.

Overall, the lower degree and smaller resource overhead of reduced
FFCC do not generally compensate for its lower intrinsic loss
tolerance, reflected by its IID threshold of approximately \(13.5\%\) compared with \(25\%\) for RHG. Under the circuit-level model, the lower graph degree narrows this disadvantage, yielding thresholds of \(1.85\%\) and \(2.75\%\), respectively, but does not eliminate it. The active-atom comparison nevertheless identifies specific low-loss resource windows in which the smaller overhead of reduced FFCC makes it favorable.

\section{Summary}

In this work, we compared periodic RHG, FFCC, and reduced FFCC in the same compound photon--atom MBQC architecture using near-deterministic photon--atom CZ gates. We developed hardware-compatible bipartite and STAP generation schemes and evaluated thresholds, LER scaling, and resource-matched LERs under a hardware-aware circuit-level loss model
with correlated bond-loss propagation. Scheme choice leaves the RHG threshold essentially unchanged, produces modest shifts for the FFCC variants, and affects the LER more clearly. Under the baseline model, RHG has the highest threshold, \(2.75\%\), and the lowest LER over most matched per-layer qubit and active-atom budgets. Reduced FFCC is favored
in specific low-loss resource windows, although its lower degree and overhead generally do not compensate for its lower IID loss tolerance.

The threshold ordering changes when selected intermodule CZ connections carry additional loss. RHG degrades more rapidly because two connections per photon carry the intermodule penalty, compared with one for reduced FFCC, so at higher intermodule loss the latter reaches the higher threshold.
Our results show that when the hardware supports the native gates and connectivity required for MBQC, the benefit of reduced graph degree
cannot be assessed in isolation, but must be considered together with
differences in intrinsic IID loss tolerance, the distribution of loss
across the generation schedule, and the associated hardware overhead.
As the hardware matures, the framework developed here for comparing codes can be extended to more detailed, architecture-specific error models and a broader range of codes. With the broader aim of increasing the tolerable loss per operation, the present work compares RHG with lower-degree codes, while future work can explore moderately higher-degree codes with higher IID-loss thresholds~\cite{nickerson2018measurement}, potentially improving per-operation loss tolerance.

\section*{Acknowledgments}
We thank David Dentelski for fruitful discussions and technical help. We acknowledge support from the Israel Innovation Authority.

\section*{Author contributions}
All authors contributed to the research, simulation development, analysis of the results, and writing of the manuscript. D.B.P. and Y.J. led the work. Generative AI tools were used to assist with code refinement and manuscript editing.


\bibliographystyle{quantum}
\bibliography{mybib}

\clearpage
\onecolumngrid
\section{Supplementary Material}

\setcounter{figure}{0}
\setcounter{table}{0}
\renewcommand{\thefigure}{S\arabic{figure}}
\renewcommand{\thetable}{S\arabic{table}}

\renewcommand{\theHfigure}{S\arabic{figure}}
\renewcommand{\theHtable}{S\arabic{table}}

\subsection{Error model and loss-aware decoding}

\label{app:error_model_decoding}

\subsubsection{Circuit-level loss model}

We use the circuit-level loss model summarized in Table~\ref{tab:error_model}. Photon loss is assumed to be the dominant error mechanism, while photonic Pauli errors, atomic loss, and intrinsic atomic errors are neglected. Photon initialization, photon participation in a photon--atom CZ gate, STAP, and photon measurement are each assigned the same effective loss probability \(p\). Atom initialization is treated as loss-free because failure is heralded and corrected before any interaction, and atoms are assumed not to be lost during CZ gates. Since atomic measurement consists of STAP followed by photon detection, it is assigned loss probability \(2p\). Loss in photonic \(H\) gates is neglected, with optical-path loss between gates assigned to the subsequent lossy operation~\cite{qs2026arxiv}. 

A photon cycle begins at photon generation or STAP and ends at photon detection. Loss is heralded only when the photon is not detected, so the measurement identifies which photon was lost but not the point along its trajectory at which the loss occurred~\cite{Baranes_2026,qs2026arxiv}.

Other hybrid emitter-based studies include photon distinguishability and spin decoherence in addition to photon loss~\cite{dessertaine2026enhancedfaulttolerancephotonicquantum,chan2025prxquantum}. These channels play a different role in our architecture. The photon--atom CZ gate is based on direct interaction and does not rely on interference between independently generated photons, so it is not subject to the same inter-photon indistinguishability errors as fusion measurements~\cite{qs2026arxiv}. Atomic dephasing is also expected to be subdominant on the timescales considered because each atom participates in only a bounded number of rapid photon cycles before STAP or measurement. By contrast, in spin-emitter fusion architectures, the spin retains quantum information through repeated photon-emission and fusion attempts, making spin decoherence an explicit noise channel~\cite{dessertaine2026enhancedfaulttolerancephotonicquantum,qs2026arxiv}.

\subsubsection{Bond-loss propagation}

We distinguish between physical photon loss and \emph{bond loss}, namely the absence of an intended CZ connection~\cite{li2010fault,auger2018fault}. If a photon is lost before completing its scheduled sequence of CZ gates, all subsequent CZ gates involving that photon fail. Consequently, physical photon loss can remove several intended bonds and modify the stabilizers and correlation surfaces of neighboring qubits. We refer to this effect as \emph{bond-loss propagation}.

To represent this process, each photon cycle is divided into intervals separated by its scheduled CZ gates. Conditioned on the photon being found lost at measurement, each interval is assigned the conditional probability that the loss occurred during that part of the trajectory~\cite{gu2024optimizingquantumerrorcorrection,Baranes_2026,qs2026arxiv}. If the loss is assigned to interval \(k\), let \(S_k\) denote the atoms involved in all CZ gates scheduled after that interval. These remaining interactions are
\begin{equation}
U_{S_k}=\prod_{a\in S_k}\mathrm{CZ}_{p,a}.
\end{equation}
Using
\begin{equation}
\mathrm{CZ}_{p,a}
=
\ket{0}\!\bra{0}_{p}\otimes I_a
+
\ket{1}\!\bra{1}_{p}\otimes Z_a,
\end{equation}
an unobserved projection of the lost photon leaves an unknown value \(b\in\{0,1\}\), resulting in the application of \((Z_{S_k})^b\) to the neighboring atoms, where
\begin{equation}
Z_{S_k}=\prod_{a\in S_k}Z_a.
\end{equation}
This results in the correlated dephasing channel 
\begin{equation}
\mathcal{E}_{S_k}(\rho)
=
\frac{1}{2}\rho
+
\frac{1}{2}Z_{S_k}\rho Z_{S_k}.
\label{eq:bond_loss_channel}
\end{equation}

Thus, the missing CZ gates generated by a single photon-loss event are represented by one correlated \(Z\)-error mechanism~\cite{yu2025processingdecodingrydbergdecay,qs2026arxiv} rather than by independent errors on the neighboring atoms as considered in~\cite{gu2025faulttolerantarchitectures,gu2024optimizingquantumerrorcorrection}.

\subsubsection{Generation-scheme dependence}

The bipartite and STAP generation schemes redistribute the same basic loss opportunities. In the bipartite scheme, a photonic qubit is generated, completes all of its CZ gates as a photon, and is then detected, while an atomic qubit undergoes STAP only at readout. In the STAP scheme, qubits transferred before completing their CZ sequence perform one or more CZ gates while atomic and the remaining gates as photons. For a degree-four qubit, this reduces the maximum number of lossy photon--atom CZ gates from four to three. It also reduces the number of possible loss intervals after the qubit becomes photonic, which may reduce the loss-time ambiguity available to the decoder.

The STAP operation does not add a net non-CZ loss opportunity: it
replaces photon generation for qubits that would be photonic in the
bipartite scheme, or moves the STAP component of atomic readout earlier for qubits that would be atomic. The tradeoff is that loss during a mid-circuit STAP can remove all subsequent bonds and induce correlated errors on neighboring atoms. Qubits transferred only at readout do not incur this additional bond-loss mechanism. These competing effects provide a qualitative explanation for the code-dependent changes in threshold and logical-\(X\) LER observed in the main text.

An additional distinction appears when the logical state probes both complementary correlation surfaces. In the bipartite scheme, the two surfaces can be supported on different physical qubit types and therefore need not have comparable LERs. For RHG, previous work~\cite{qs2026arxiv} found the photon-supported correlation surface to have a substantially higher threshold than the
atom-supported surface. Consequently, the logical-\(Y\) LER closely follows the atom-supported logical-\(X\) LER because the photon-supported contribution is negligible. In the STAP scheme, this fixed physical-support asymmetry is absent because qubits change physical type during the schedule. Both logical components may therefore contribute appreciably to the LER of a general logical state. Accordingly, we expect the LER for a general logical state in the STAP scheme to approach approximately twice the logical-\(X\) LER if the two logical components contribute comparably, whereas in the bipartite scheme the general-state and logical-\(X\) LERs are expected to remain similar.

\subsubsection{Simulation and decoding}

We implement our error model using a two-level Monte Carlo procedure. For each physical loss probability \(p\), we first sample \(N\) heralded loss configurations according to the chosen code and generation scheme. Each configuration specifies which photons are absent at measurement. For every sampled configuration, we construct the corresponding loss-conditioned detector error model and perform \(M\) detector-sampling and decoding iterations. The values of \(N\) and \(M\) used for each data set are given in the corresponding figure captions~\cite{gu2024optimizingquantumerrorcorrection, qs2026arxiv}. 

For every lost photon, the possible loss intervals are mutually exclusive. We encode them in \texttt{Stim} using a sequence of \texttt{CORRELATED\_ERROR} and \texttt{ELSE\_CORRELATED\_ERROR} instructions~\cite{gidney2021stim}. Their probabilities are chosen so that the resulting distribution reproduces the conditional loss-time probabilities. The selected interval determines the correlated \(Z\) error of Eq.~\eqref{eq:bond_loss_channel}, which is propagated to the detector outcomes and logical observable. 

After this conditional replacement, the noise model is expressed entirely through Pauli error mechanisms. The correlated mechanisms are decomposed into graphlike components before standard MWPM decoding. Detector outcomes are sampled using \texttt{Stim}, and the resulting syndromes are decoded by minimum-weight perfect matching using PyMatching~\cite{higgott2021pymatchingpythonpackagedecoding, Higgott2025sparseblossom}. A shot is counted as a logical failure when the decoder's inferred logical correction disagrees with the sampled logical observable. Thresholds are estimated from finite-size crossing points, and the LER is obtained by averaging over both the sampled loss configurations and the conditional decoding iterations.

\subsection{Supporting numerical results}

\begin{figure}
\centering

\begin{minipage}[t]{0.46\textwidth}
    \centering
    \vspace{0pt}
    \stackinset{l}{0pt}{t}{2pt}{\textbf{(a)}}{%
        \includegraphics[width=\linewidth]{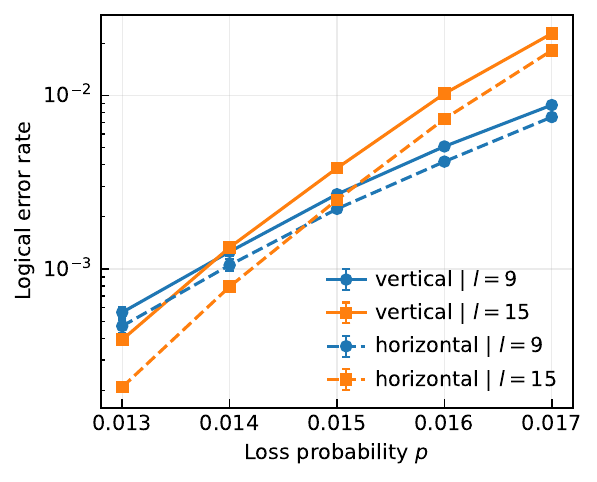}%
    }%
\end{minipage}
\hfill
\begin{minipage}[t]{0.46\textwidth}
    \centering
    \vspace{0pt}
    \stackinset{l}{0pt}{t}{2pt}{\textbf{(b)}}{%
        \includegraphics[width=\linewidth]{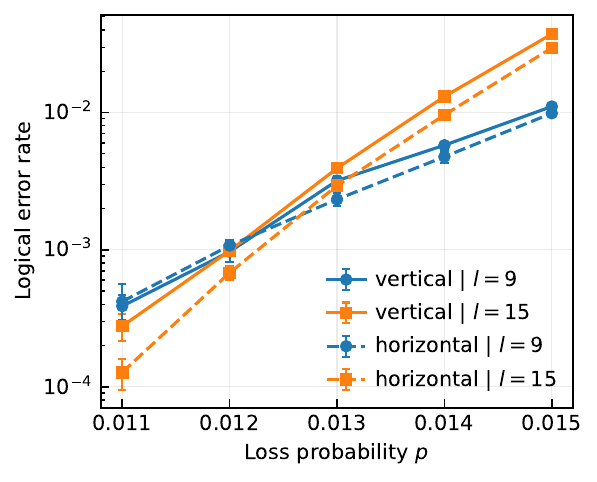}%
    }%
\end{minipage}

\caption{Circuit-level logical-\(X\) memory threshold simulations for the FFCC code with odd \(l\). 
Panel (a) shows the bipartite generation scheme, and panel (b) shows the STAP generation scheme. 
These results complement the even-\(l\) results shown in Fig.~\ref{fig:thresholds_ffcc}.}
\label{fig:thresholds_ffcc_odd}
\end{figure}

\begin{figure}
\centering

\begin{minipage}[t]{0.46\textwidth}
    \centering
    \vspace{0pt}
    \stackinset{l}{-5pt}{t}{2pt}{\textbf{(a)}}{%
        \includegraphics[width=\linewidth]{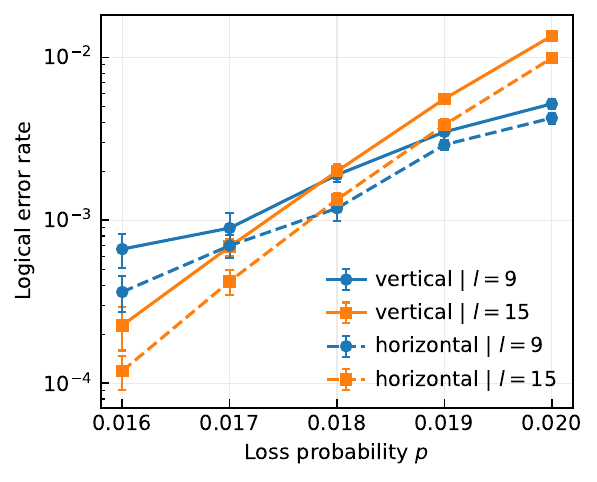}%
    }%
\end{minipage}
\hfill
\begin{minipage}[t]{0.46\textwidth}
    \centering
    \vspace{0pt}
    \stackinset{l}{0pt}{t}{2pt}{\textbf{(b)}}{%
        \includegraphics[width=\linewidth]{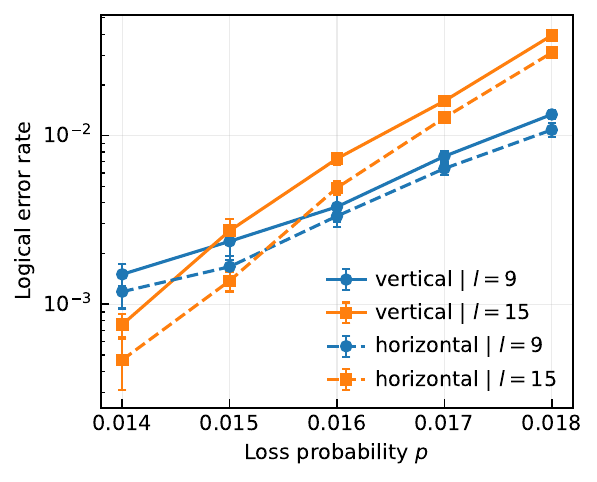}%
    }%
\end{minipage}

\caption{Circuit-level logical-\(X\) memory threshold simulations for the reduced FFCC code with odd \(l\). 
Panel (a) shows the bipartite generation scheme, and panel (b) shows the STAP generation scheme. 
These results complement the even-\(l\) results shown in Fig.~\ref{fig:thresholds_reduced_ffcc}.}
\label{fig:thresholds_reduced_ffcc_odd}
\end{figure}

\begin{figure}
\centering
\begin{minipage}[t]{0.46\textwidth}
    \centering
    \vspace{0pt}
    \stackinset{l}{-5pt}{t}{2pt}{\textbf{(a)}}{%
        \includegraphics[width=\linewidth]{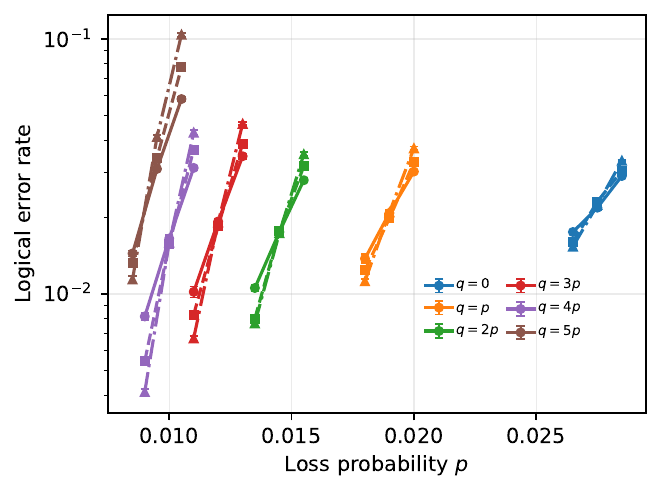}%
    }%
\end{minipage}
\hfill
\begin{minipage}[t]{0.46\textwidth}
    \centering
    \vspace{0pt}
    \stackinset{l}{0pt}{t}{2pt}{\textbf{(b)}}{%
        \includegraphics[width=\linewidth]{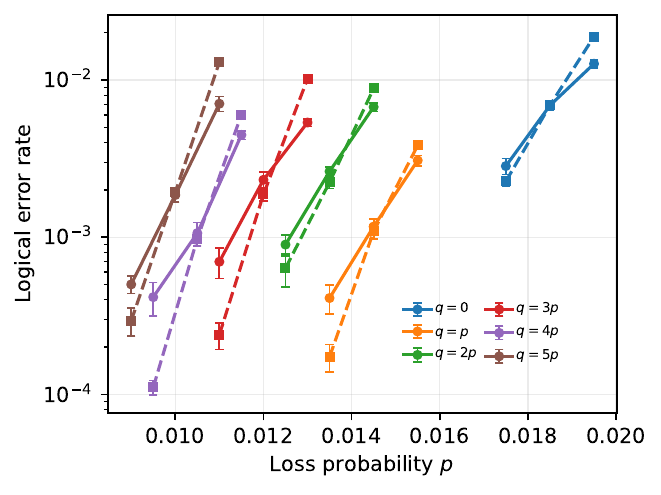}%
    }%
\end{minipage}
\caption{
Finite-distance LER curves used to extract the thresholds shown in main-text Fig.~\ref{fig:intermodule_loss_comparison}.
Intermodule CZ connections have loss probability $p_{\mathrm{inter}} = p+q$ instead of $p$.
For RHG, this applies to the third and fourth CZ connections, whereas for reduced FFCC it applies only to the third CZ connection.
(a) RHG logical-$X$ LER for $l=7,9,11$.
(b) Vertical reduced FFCC logical LER for $l=12,18$.
Each point uses $3{,}000$ loss realizations and $10{,}000$ decoding iterations per realization.
Error bars show SEM.
}
\label{fig:intermodule_loss_ler}
\end{figure}

Figures~\ref{fig:thresholds_ffcc_odd} and~\ref{fig:thresholds_reduced_ffcc_odd} show the circuit-level thresholds for FFCC and reduced FFCC with odd \(l\). These results complement the even-\(l\) data used in the main-text threshold comparison, where even \(l\) gives the higher threshold because it corresponds to a square layer geometry. 
Supplementary Fig.~\ref{fig:intermodule_loss_ler} shows the finite-distance LER crossings used to extract the RHG and reduced FFCC thresholds under additional intermodule loss reported in main-text Fig.~\ref{fig:intermodule_loss_comparison}.

\end{document}